\documentclass[12pt]{article}
\usepackage[margin=1in]{geometry}
\usepackage{amsmath,amssymb,amsthm,mathtools}
\usepackage{graphicx}
\usepackage{booktabs}
\usepackage{multirow}
\usepackage{microtype}
\usepackage{enumitem}
\usepackage{algorithm}
\usepackage{algpseudocode}
\usepackage[colorlinks=true,allcolors=blue]{hyperref}
\usepackage[nameinlink]{cleveref}

\theoremstyle{definition}

\theoremstyle{remark}

\title{Surrogate-based Bayesian calibration methods for chaotic systems: a comparison
of traditional and non-traditional approaches \thanks{This material is based upon work supported by the U.S. Department of Energy, Office of
Science, Office of Advanced Scientific Computing and Sandia National Laboratories Advanced Science and Technology Laboratory Directed Research and Development program.
Sandia National Laboratories is a multimission laboratory managed and operated by National Technology and Engineering Solutions of Sandia, LLC., a wholly owned subsidiary of Honeywell International, Inc., for the U.S. Department of Energy’s National Nuclear Security Administration under contract DE-NA0003525. This paper describes objective technical results and analysis. Any subjective views or opinions that might be expressed
in the paper do not necessarily represent the views of the U.S. Department of Energy or the United States Government.
This research used resources of the National Energy Research Scientific Computing Center (NERSC), a U.S. Department of Energy Office of Science User Facility located at Lawrence Berkeley National Laboratory, operated under Contract No. DE-AC02-05CH11231
using NERSC award DOE-ERCAPm3876. This research was supported in part through computational resources and services provided by Advanced Research Computing at the University of Michigan, Ann Arbor.
The authors report there are no competing interests to declare.}}
\author{
Maike F.\ Holthuijzen\thanks{Sandia National Laboratories ( \href{mailto:mfholth@sandia.gov}{mfholth@sandia.gov}, \href{mailto:ekrath@sandia.gov}{ekrath@sandia.gov}, \href{mailto:tacatan@sandia.gov}{tacatan@sandia.gov} )}
\and Atlanta Chakraborty\thanks{University of Michigan, Dept.\ of Mechanical Engineering, Ann Arbor, MI 48109 ( \href{mailto:atlantac@umich.edu}{atlantac@umich.edu} )}
\and Elizabeth Krath\footnotemark[2]
\and Tommie Catanach\footnotemark[2]
}
\date{}

\begin{document}
\maketitle

\begin{abstract}
Parameter calibration is essential for reducing uncertainty and improving predictive fidelity in physics-based models, yet it is often limited by the high computational cost of model evaluations. Bayesian calibration methods provide a principled framework for combining prior information with data while rigorously quantifying uncertainty. In this work, we compare four emulator-based Bayesian calibration strategies—Calibrate–Emulate–Sample (CES), History Matching (HM), Bayesian Optimal Experimental Design (BOED), and a goal-oriented extension of BOED (GBOED)—using the Lorenz ’96 multiscale system and a two-layer quasi-geostrophic (QG) model as test cases. The proposed GBOED formulation explicitly targets information gain with respect to the calibration posterior, aligning design decisions with downstream inference. We assess methods using accuracy and uncertainty quantification metrics, convergence behavior under increasing computational budgets, and practical considerations such as implementation complexity and robustness.

For the Lorenz ’96 system, CES, HM, and GBOED all yield strong calibration performance, even with limited numbers of model evaluations, while standard BOED generally underperforms in this setting. Differences among the strongest methods are modest, particularly as computational budgets increase. For the QG system, all methods produce reasonable posterior estimates, and convergence behavior is more consistent, reflecting smoother parameter–statistic relationships and improved identifiability. Overall, our results indicate that multiple emulator-based calibration strategies can perform comparably well when applied appropriately, with method selection often guided more by computational and practical considerations than by accuracy alone. These findings highlight both the limitations of standard BOED for calibration and the promise of goal-oriented and iterative approaches for efficient Bayesian inference in complex dynamical systems.
\end{abstract}


\section{Introduction}
\label{sec:intro}
Uncertainty quantification (UQ) is essential for validating predictive models in physics-based systems, where parameter choices propagate through nonlinear, often chaotic dynamics. Calibration, the process of inferring parameters from data with quantified uncertainty, is a many-query problem: it requires repeated evaluations of expensive simulators and thus demands methods that are both statistically reliable and computationally efficient.

A common strategy is to replace the simulator with an \emph{emulator}, such as a Gaussian process (GP), that provides fast predictions with uncertainty. Emulator accuracy in calibration hinges on how simulation runs are chosen: informative design points can dramatically reduce cost, whereas uninformed sampling wastes evaluations and degrades posterior inferences.

We study Bayesian calibration for chaotic dynamical systems using emulators and compare four data-acquisition paradigms: Bayesian Optimal Experimental Design (BOED) \cite{lindley1956measure,chaloner1995bayesian,ginebra2007measure} and its goal-oriented extension (GBOED) \cite{bernardo1979expected,attia2018goal,zhong2024goal}, Calibrate–Emulate–Sample (CES) \cite{cleary2021calibrate}, and History Matching (HM) \cite{craig1997pressure}. While BOED has a long history in physical experimentation and sensor placement \cite{wu2021efficient,ricciardi2024bayesian,ferrolino2020optimal}, its use for selecting \emph{computer simulations} targeted to emulator-based calibration remains comparatively underexplored.

\paragraph{Contributions}
We make the following contributions:
\begin{enumerate}
  \item We introduce a goal-oriented BOED (GBOED) algorithm that directly targets information gain about the \emph{calibration posterior}, rather than about the emulator alone, making the design criterion explicitly aligned with downstream inference.
  \item We provide a comparative evaluation of BOED, GBOED, CES, and HM on the multiscale Lorenz'96 multiscale system, quantifying trade-offs among computational cost, implementation complexity, and posterior accuracy/uncertainty.
  \item We assess the most promising methods on a computationally demanding quasi-geostrophic (QG) model, demonstrating performance and practical considerations in a higher-fidelity fluid-dynamics setting.
\end{enumerate}
Results from applying the main methods to the two systems clarify when BOED and GBOED-based methods offer advantages for calibration, and how they compare to CES and HM when emulators drive the final MCMC-based inference.

\subsection{Background and motivation}
All methods considered in this study fall under the category of Bayesian calibration, in which parameter estimates and their uncertainty are rigorously quantified via a probability distribution, rather than a fixed value. From the Bayesian inverse-problems viewpoint, ensemble-based samplers provide optimization-free routes to posterior exploration. The \emph{ensemble Kalman sampler} (EKS) interprets ensemble updates as a gradient flow for a mean-field approximation to the posterior \cite{garbuno2020interacting}, while \emph{ensemble Kalman inversion} (EKI) adapts Ensemble Kalman Filter ideas to static inverse problems and has been analyzed/extended for nonlinear forward maps and regularization \cite{iglesias2013ensemble, kovachki2019ensemble,tienstra2024early, chada2020iterative, huang2022iterated}. We are most interested in \emph{emulator-driven calibration}, in which a surrogate (often a Gaussian process model) replaces an expensive model   \cite{kennedy2001bayesian,wilkinson2010bayesian,gramacy2020surrogates}. Emulators enable the thousands of forward evaluations required by uncertainty quantification (UQ) procedures such as Markov chain Monte Carlo (MCMC), while keeping total compute manageable. In some inference settings, emulators are trained in an amortized fashion, with the goal of supporting many downstream inference tasks across varying observational data \cite{zammit2025neural}. However, calibration to a fixed observational data set does not fall into this category: accuracy is only required in regions of parameter space consistent with the observed data. This distinction motivates adaptive, calibration-aware design strategies that concentrate simulation effort where it is most informative for posterior inference, rather than seeking globally accurate surrogates. Since the seminal paper that introduced the formalism of Bayesian calibration of computational models in 2001 \cite{kennedy2001bayesian}, interest in this method has grown steadily in several fields \cite{viana2021survey}, including biology \cite{helleckes2022bayesian}, engineering \cite{durr2023bayesian, ehrett2021simultaneous}, physics \cite{ling2013challenging} and climate science \cite{jackson2008error, dunbar2021calibration, chang2014fast, williamson2015identifying, king2024bayesian, yarger2024autocalibration, smith2024fair}.

Here, we compare the performance of four emulator-based calibration methods (CES, HM, BOED, GBOED) on calibrating two chaotic dynamical systems: the Lorenz '96 multiscale system and the quasi-geostrophic (QG) system. We focus on systems where the forward model is moderately expensive and exhibits sensitive dependence on parameters and initial conditions, a combination that makes calibration challenging. We specifically study how different design and calibration choices interact in the regions of parameter space that matter most for inference.

The design challenge in Bayesian calibration refers to selecting the limited set of model runs needed to train the emulator while maximizing its accuracy for calibration tasks. Traditional sampling strategies, such as Monte Carlo sampling and space-filling designs like Quasi Monte Carlo \cite{owen2003quasi} and Latin hypercube sampling \cite{mckay+bc79} improve efficiency compared to dense grids but are not specifically tailored to the downstream goal of calibration. Methods such as History Matching (HM) \cite{williamson2013history} and Calibrate, Emulate, Sample (CES) \cite{cleary2021calibrate} aim to address this gap by focusing sampling on regions of the parameter space relevant to calibration. HM iteratively refines the parameter space by eliminating implausible regions, dramatically reducing computational costs \cite{williamson2013history}. HM does not provide UQ on its own, although it can be combined with MCMC sampling to produce a posterior distribution over calibration parameters, which is the approach we take. CES is unique in that it combines a Kalman filtering/sampling step, GP emulation, and MCMC sampling to iteratively refine the posterior, providing systematic UQ at each stage \cite{cleary2021calibrate}.

Leveraging the formalism of Bayesian theory, BOED and GBOED extend these ideas by framing the design of model evaluations as an optimization problem. Within its standard formulation, BOED selects simulation designs that maximize the expected information gained about the emulator, leveraging the probabilistic nature of GPs to guide decisions. GBOED refines this approach further by focusing on regions of the parameter space that directly reduce uncertainty in the posterior distribution, making it particularly well-suited for calibration tasks. In essence, GBOED works by resolving intermediate calibration problems, leveraging the surrogate and its associated uncertainty. While BOED has been shown to achieve reliable calibration results for expensive models with fewer model evaluations \cite{dunbar2022ensemble}, the use of GBOED for calibration-specific tasks has not been explored in the literature.

This paper is structured as follows: first, we review GP emulators (Section \ref{sec:CalibwEmulators}) and  provide an overview of the main methods (BOED, GBOED, CES, and HM) (Sections \ref{sec:boed}-\ref{sec:HM}). Next, we describe the experimental setups for the Lorenz '96 (Section \ref{sec:Experiment}), validation (Section \ref{sec:Validation}), and results (Section \ref{sec:results}). We describe the quasi-geostrophic system and results in Section \ref{sec:QGsystem}. Finally, we provide a discussion and concluding remarks in Section \ref{sec:conclusions}.

\section{Methods}
\label{sec:Methods}
In this study, we assume we have access to simulations from a model $\mathcal{G}$, which is typically computationally expensive. Due to the computational intensity of $\mathcal{G}$ we seek to approximate it with a cheap emulator whenever we need to evaluate it many times for the purpose of model calibration. Typically the emulator does not capture the full output of $\mathcal{G}$, e.g. a very high dimensional spatiotemporal dataset, but only a reduced dimensional output comprised of important quantities of interest for solving the calibration problem. Since output quantities of interest from $\mathcal{G}$ are usually $p>1$ dimensional, two types of emulators may be considered. If outputs can be reasonably considered independent of one another, $p$ individual emulators can be used for modeling each component output. Alternatively, a multi-output emulator can be used to model correlation among the $p$ component outputs. Here, we consider independent GP emulators for each output dimension for reduced computational burden (see SM5 for a description of the decorrelation procedure applied to outputs prior to GP fitting). Gaussian processes (GPs) are a natural choice for emulation due to their well-calibrated uncertainty quantification (UQ), flexibility, and analytical tractability, making them particularly well-suited for Bayesian calibration \cite{williams2006gaussian}. However, for larger dimensional problems, GP approximations may be required; scalable GP approximations include techniques such as compactly supported kernels \cite{gneiting2002compactly, kaufman2011efficient}, fixed-rank kriging \cite{cressie2008fixed} inducing points \cite{banerjee2008gaussian, quinonero2005unifying, hensman2013gaussian} local divide-and-conquer \cite{emery2009kriging,gramacy2016lagp} and the Vecchia/nearest-neighbor approximation 
\cite{datta2016hierarchical, vecchia1988estimation,katzfuss2021general, katzfuss2020vecchia, cao2022scalable}.
For larger problems, alternative emulators can be used, including deep neural networks \cite{bocquet23, lutsko+cgmm21}, polynomial chaos expansions \cite{novak2018polynomial}, and other machine learning methods \cite{weber+chkl20}. Additionally, in applications such as full climate fields, the number of raw outputs can be enormous, and fitting independent GPs to each component is neither computationally nor statistically sensible. One standard approach is to work in a reduced output space: compute a basis from an ensemble of simulator runs (using principal components analysis  or other low-rank decompositions), retain $k$ coefficients that explain most variability, and emulate only those coefficients while treating the truncation residual as additional noise.

\subsection{Bayesian Calibration with GP emulators}
\label{sec:CalibwEmulators}
Bayesian calibration provides a probabilistic framework for inferring unknown model parameters and incorporating both prior knowledge and observational data. Given observations $y_{obs}$
and a computational model $\mathcal{G}$ that maps parameters $\theta$ to model output, we assume the relationship $y_{obs} = \mathcal{G}(\theta) + \eta$ where $\eta$ accounts for observational noise and model discrepancy. The goal is to estimate the posterior distribution of $\theta$ given $y_{obs}$, which follows from Bayes’ Theorem:

\begin{equation}\label{eq:simplelikelihood}
p(\theta \mid y_{obs}) = \frac{p(y_{obs} \mid \theta) p(\theta) }{p(y_{obs})}.
\end{equation}

\noindent When we assume that the likelihood, $p(y_{obs} \mid \theta)$, is Gaussian, then

\begin{equation}\label{eq:likelihoodmain}
p(y_{obs} \mid \theta) = \frac{1}{(2\pi)^{p/2} |\Gamma_{obs}|^{1/2}} \exp \left (-\frac{1}{2} (y_{obs} - \mathcal{G}(\theta))^T \Gamma_{obs}^{-1}(y_{obs} - \mathcal{G}(\theta)) \right ),
\end{equation}
 where $\Gamma_{obs}$ reflects uncertainty among observations and $p$ is the output dimension. In principle, we can find the posterior using MCMC to generate samples from the distribution, but this would be computationally prohibitive, because it would require evaluating $\mathcal{G}$ thousands of times. Instead, we consider a GP emulator that approximates the mapping from $\theta$ to $\mathcal{G}(\theta)$. While $\theta$ denotes the model parameters to be calibrated, we use $\mathbf{X} =\left \{x_1^\top, x_2^\top, \dots x_m^\top \right \}$ to represent inputs to the GP emulator in order to distinguish calibrating the model from training the emulator. These inputs correspond to sampled locations in the parameter space, which may include $\theta$.

Here, we provide a brief review of GPs for surrogate modeling. For more details see \cite{williams2006gaussian} or \cite{gramacy2020surrogates}. Consider the expensive computer model $\mathcal{G}$ represented as a function $f: \mathbb{R}^d \rightarrow \mathbb{R}^p$ that maps inputs to outputs, whose pairs over $n$ simulations comprise data $D = (\mathbf{X}, \mathcal{G}(\mathbf{X}))$, where inputs (model parameters) are $\mathbf{X} = \{x_1^\top, x_2^\top, \dots x_m^\top \} \in \mathbb{R}^{m \times d}$ and 
$\mathcal{G}(\mathbf{X}) = \{\mathcal{G}(x_1)^\top, \mathcal{G}(x_2)^\top, \dots\penalty0  \mathcal{G}(x_m)^\top \}  \in \nobreak \mathbb{R}^{m \times p}$
are corresponding model outputs. Since we use independent GP emulators for each output dimension, we consider modeling the individual components $\mathcal{G}_i(\mathbf{X}), i = 1\dots p$. For notational simplicity, let $\mathbf{Y}$ represent a single component output of $\mathcal{G}(\mathbf{X})$. Assuming a zero-mean GP prior for each component output of $f$ implies that $\mathbf{Y}$ follows a multivariate normal distribution (MVN) with a mean of zero and and a covariance $\Gamma_{GP}$ that is determined by inverse distances between inputs $\mathbf{X}$:

\begin{equation}\label{eq:Ydistmain}
\mathbf{Y} \sim \mathcal{N}(\mathbf{0}, \Gamma_{GP}(\mathbf{X} ))
\quad \mbox{where, e.g.,} \quad \Gamma_{GP}^{k \ell} = \tau^2(k(q(x_k, x_{\ell})) + g  \mathbb{I}_{\{k=\ell \}})),
\end{equation}
where a kernel, $k(\cdot)$, transforms distances, and $q(\cdot, \cdot)$ calculates a scaled squared Euclidean distance:

\begin{equation*}
q(x_i, x_j) =   \left ( \sum_{\ell = 1}^d   
\frac{|| x_{i,\ell} - x_{j,\ell}||^2}{\gamma_{\ell}} \right )^{1/2}.
\end{equation*}
The $\boldsymbol{\gamma} = (\gamma_1, \gamma_2, \dots \gamma_{d})$ are {\em lengthscale} or {\em range} parameters, where $1 / \gamma_{\ell}$ determines the ``relevance'' of the $x_{\ell}$ input,  $\tau^2$ governs the magnitude of the response, and the nugget $g$ divides the magnitude between signal $(\tau^2)$ and noise $(\tau^2 g)$. In this study, we use the Mat\'ern kernel with smoothness level fixed at 3.5. This implies translation-invariant covariance (dependence only on separation). Other non-stationary kernels with input-dependent length-scales could be used if stationarity is inappropriate for a given application, but in this study we assume stationarity. The GP model in Eq. (\ref{eq:Ydistmain}) for $\mathbf{Y}$ gives rise to a MVN likelihood:
\begin{equation}\label{eq:likelihoodY}
L(\boldsymbol{\phi} ; D) \propto |\Gamma_{GP}|^{-1/2} \mathrm{exp} \left( -\frac{1}{2} \mathbf{Y}^\top \Gamma_{GP}^{-1} \mathbf{Y} \right).
\end{equation}
We express the framework in Eq.~(\ref{eq:Ydistmain}) as $\mathcal{GP} (D)$, which is defined by a choice of $k(\cdot)$, $q(\cdot)$, and hyperparameters $\boldsymbol{\phi} = (\boldsymbol{\gamma}, \tau^2, g)$. Hyperparameter estimates $\hat{\boldsymbol{\phi}}$ are learned from the data, $D$, via the likelihood in \ref{eq:likelihoodY} via maximum a posteriori (MAP) estimation. Typically, the predictive mean from $\mathcal{GP}(D)$ at new locations $\mathcal{X}$ is of primary interest. 
Ultimately, $\mathcal{GP}(D)$ can be used to obtain the posterior predictive distribution $\mathbf{Y}(\mathcal{X})| D$, which is conditional on estimated hyperparameter values. The resulting posterior predictive distribution, also an MVN, can be obtained through the MVN conditioning equations: $\mathbf{Y}(\mathcal{X}) | D \sim \mathcal{N} (\mu_{GP}(\mathcal{X}), \Gamma_{GP}(\mathcal{X}))$.  Therefore, incorporating the GP predictive mean and variance into Eq. (\ref{eq:likelihoodmain}) provides the following likelihood for the emulator-based calibration problem:

\begin{equation}\label{eq:likelihoodwGP}
p(y_{obs} \mid \theta, \mathcal{GP}(D)) = \frac{ e^{-\frac{1}{2} (y_{obs} - \mu_{GP}(\theta))^T (\Gamma_{GP}(\theta) + \Gamma_{obs})^{-1}(y_{obs} - \mu_{GP}(\theta)) }}{(2\pi)^{d/2} |\Gamma_{GP}(\theta)+\Gamma_{obs}|^{1/2}}.
\end{equation}

Because obtaining predictions from the GP emulator at any new inputs is cheap, MCMC can be used to perform Bayesian calibration via Eq. (\ref{eq:likelihoodwGP}) while incorporating the model discrepancy due to the use of a GP emulator. For each of the methods considered in this paper, the final posterior for $\theta$ is computed via an MCMC algorithm using the likelihood in Eq. (\ref{eq:likelihoodwGP}). The key challenge, however, in using GPs as part of a calibration method is judiciously choosing simulation training data $D$ to produce the most accurate and efficient emulator to be used in the calibration problem. 

\subsection{Bayesian Optimal Experimental Design}
\label{sec:boed}
Optimal Experimental Design (OED) focuses on identifying and optimizing data collection strategies to maximize the amount of information gained from experiments. In this context, information gain typically refers to changes in beliefs toward a reduction in the uncertainty of model parameters. Within the Bayesian paradigm, BOED seeks to collect data that maximizes the Expected Information Gain (EIG), which quantifies how much the posterior distribution reduces uncertainty relative to the prior, measured using Kullback-Leibler (KL) divergence. BOED is a theoretically rigorous approach but can be computationally intensive and sensitive to assumptions \cite{chaloner1995bayesian}. In this context, we emphasize that the data being collected are not direct observations ($y_{obs}$) but rather data arising from simulations used to fit an emulator. 

Ideally, we formulate the BOED problem to maximize the EIG about the model parameters we wish to infer. Suppose we have initial data $D = (\mathbf{X}_{init}, \mathbf{Y}_{init})$ and that a zero-mean GP is fit to $D$, e.g. ($\mathcal{GP}(D)$). Let $\mathbf{X}$ represent an initial design of inputs chosen via a space-filling method such as Latin Hypercube Sampling (LHS) or Quasi Monte Carlo sampling. These inputs are selected to provide a representative coverage of the parameter space and to initialize the GP emulator. We can measure how much information we expect to gain about $\theta$ when training on $D$ by computing the EIG:

\begin{align}\label{eq:EIGxstar}
\textup{EIG}(\mathbf{X}) &= \int p(\mathbf{Y} | \mathbf{X}) \int p(\theta|y_{obs}, \mathcal{GP}(D)) \log \frac{p(\theta|y_{obs}, \mathcal{GP}(D))}{p(\theta)} d\theta d \mathbf{Y} \nonumber \\
&= \int p(\mathbf{Y} | \mathbf{X}) \int p(\theta|y_{obs}, \mathcal{GP}(D)) \log \frac{p(y_{obs} | \theta, \mathcal{GP}(D))}{p(y_{obs} | \mathcal{GP}(D))} d\theta d \mathbf{Y}
\end{align}
 In Eq. (\ref{eq:EIGxstar}), $\mathbf{Y}|\mathbf{X}$ are realizations from the GP prior. In addition, $p(\theta|y_{obs}, \mathcal{GP}(D))$ is the posterior distribution of parameters $\theta$, $p(\theta)$ is the prior distribution on $\theta$, and $p(y_{obs} | \theta, \mathcal{GP}(D))$ is the likelihood. We wish to choose the design $\mathbf{X}^*$ that maximizes the EIG; that is, given bounds on the input space ($\Xi$), we wish to find
$\mathbf{X}^* = \arg \max_{\mathbf{X} \in \Xi} \textup{EIG}(\mathbf{X})$. The expression in \ref{eq:EIGxstar} represents the ``idealized'' OED problem and is rarely implemented in practice. Computing the EIG in the idealized OED problem is challenging, because for each candidate design $\mathbf{X}$, we must sample realizations $\mathbf{Y}$ from the GP prior predictive distribution, solve the calibration problem to obtain $p(\theta|y_{obs}, \mathcal{GP}(D))$, and estimate the KL divergence or the marginal model evidence  $p(y_{obs} | \mathcal{GP}(D))$. 

To solve the idealized OED problem in Eq. (\ref{eq:EIGxstar}), a sequential approach may be taken, where we iteratively optimize the EIG to determine an initial design $\mathbf{X}_0$, run simulations, and then conditionally optimize additional simulations. For the updated training dataset, we include new points $\mathbf{X}_1 = {\mathbf{X}_0 \cup \mathbf{X}_{new}}$, where $\mathbf{X}_{new}$ represents a newly selected batch of inputs, which can be run as an ensemble in parallel. This iterative process ensures that simulation samples are strategically placed to improve calibration performance of the emulator, rather than being driven by uninformed priors. Without this sequential refinement, samples are less effective, as they rely solely on prior information that may be uninformative. However, sequential refinement is computationally expensive, as it requires repeatedly solving the calibration problem and updating the emulator. In what follows, we emphasize two practical experimental design strategies: standard BOED, which maximizes EIG about the emulator, and goal-oriented BOED (GBOED), which prioritizes information gain about the inferred parameters. For completeness, we describe the ``idealized'' BOED formulation as a conceptual gold standard (though computationally infeasible for real-world problems) and compare it against BOED and GBOED in SM3.

When applying BOED to maximize the EIG for a GP emulator, we aim to maximize information gain about GP outputs, which is directly tractable. Assuming the specification of a GP prior for $\mathbf{Y}$, consider an extended candidate set of inputs $\mathbf{X^\prime}$ where $\mathbf{X} \subset \mathbf{X^\prime}$ and for which $\mathbf{Y}' \sim MVN(\mathbf{0}, \mathbf{K}(\mathbf{X^\prime}))$ reflects the GP prior evaluated at $\mathbf{X}'$. The posterior predictive distribution of $\mathbf{Y}'$ at inputs $\mathbf{X^\prime}$ given training data $D = (\mathbf{X}, \mathbf{Y})$ is denoted as $\mathbf{Y}' | \mathbf{X^\prime}, \mathcal{GP}(D)$. In the standard BOED approach, the EIG is defined as Eq. (\ref{eq:EIG27}):
\setlength{\belowdisplayskip}{1em} 

\begin{equation}\label{eq:EIG27}
\textup{EIG}(\mathbf{X})= \int p(\mathbf{Y} | \mathbf{X}) \int p(\mathbf{Y}^\prime |\mathbf{X^\prime} , \mathcal{GP}(\mathbf{X}, \mathbf{Y})) \log \frac{p(\mathbf{Y}^\prime| \mathbf{X^\prime} , \mathcal{GP}(\mathbf{X}, \mathbf{Y}))}{p(\mathbf{Y}^\prime|
\mathbf{X}')} \, d\mathbf{Y^\prime} \, d \mathbf{Y}. \end{equation} 
In Eq. \ref{eq:EIG27}, $\mathbf{X}^\prime$ 
is a fixed set of test locations; during each experiment optimization stage we fix $\mathbf{X}^\prime$ (hence we abbreviate to $\textup{EIG}(\mathbf{X})$ below). The inner integral marginalizes the random vector $\mathbf{Y}'$, and the outer integral averages over possible training responses $\mathbf{Y}$ drawn from the GP prior predictive at $\mathbf{X}$. The expression quantifies how much we expect the GP posterior to differ from the GP prior at candidate locations $\mathbf{X}'$, on average. Let the (zero-mean) GP prior have covariance matrix $\mathbf{K}_{X'X'}$, and let the posterior predictive variance at $\mathbf{X}'$ \emph{after} observing data $\mathcal{D} = (\mathbf{X}, \mathbf{Y})$ be denoted by $\mathbf{\Sigma}'$. Then, the EIG in (\ref{eq:EIG27}) can be expressed in closed form as the expected KL divergence (over $\mathbf{Y}$) between two multivariate Gaussian distributions (Eq. \ref{eq:EIG_closed}) (see proof in SM4):

\begin{align}\label{eq:EIG_closed}
EIG(\mathbf{X}) &= \frac{1}{2}  \log\left(\frac{\det(\mathbf{K}_{X^\prime, X^\prime}) }{\det(\mathbf{\Sigma}^\prime)}\right) ,\\
\mathbf{\Sigma}^\prime &= \mathbf{K}_{X^\prime, X^\prime} - \mathbf{K}_{X^\prime, X} \mathbf{K}_{X, X}^{-1} \mathbf{K}_{X, X^\prime}. \nonumber
\end{align}
The algorithm for standard BOED is given in Algorithm 2.1.

Not all emulator predictions are equally relevant to calibration. Let $m(\theta)$ and $\Sigma_{total}(\theta)$ denote the emulator mean and total covariance (including $\Gamma_{obs}$). Locally, the calibration information carried by the outputs is governed by how $m(\theta)$ changes with $\theta$ relative to $\Sigma_{total}(\theta)$. Outputs that are weakly sensitive to $\theta$ or highly noise-dominated contribute little. Thus, choosing new simulations solely to reduce emulator error everywhere in $\theta$-space can be misaligned with calibration. To address this, we introduce a goal-oriented BOED (GBOED) strategy, which focuses on predicting values most relevant to the calibration posterior. Instead of choosing a generic set of test inputs $\mathbf{X}'$, GBOED utilizes test locations drawn from the current approximated posterior of $\theta$, i.e., $\mathbf{X}' \sim p(\theta | y_{obs}, \mathcal{GP}(D))$, thereby aligning the training design more closely with the calibration task (see Algorithm 2.2). While BOED and GBOED differ in the choice of $\mathbf{X}'$, we note that if $\mathbf{X}'$ becomes dense in the input space (i.e., fully explores the emulator domain), then GBOED recovers standard BOED. We prove this in the finite case (see SM1). This theoretical equivalence justifies GBOED as a goal-oriented simplification of BOED that prioritizes the most relevant parts of the input space for calibration. In principle, emulator miscalibration could bias the posterior toward regions where the GP is less accurate. However, in practice, each $\mathbf{X}_{new}$ is appended to the training set and the GP is refit, which reduces predictive error locally. Throughout, ``standard BOED" denotes the closed-form emulator-EIG in Eq. (2.7) evaluated at a fixed $\mathbf{X}'$. GBOED uses the same objective but sets $\mathbf{X}' \sim p(\theta | y_{obs}, \mathcal{GP}(D))$ to target calibration, whereas Eq. (2.6) is the idealized parameter-EIG we include for context. For the interested reader, SM4 shows the differences in BOED, GBOED and idealized OED when applied to a toy calibration problem.

\begin{algorithm}[H]
\caption{Standard Bayesian Optimal Experimental Design (BOED)}
\label{alg:standard_boed}

\textbf{Input:} Bounds for input space $\Xi$, initial number of samples $n_0$, batch size $B$, total simulation budget $N$, GP emulator \texttt{GP} \\
\textbf{Output:} Samples from calibrated posterior distribution $p(\theta \mid y_{\text{obs}}, \mathcal{GP}(D))$

\begin{algorithmic}[1]
\State Select an initial set of $n_0$ input samples $\mathbf{X}_{\text{init}}$ from $\Xi$.
\State Run the simulation $\mathcal{G}$ for each input in $\mathbf{X}_{\text{init}}$ to generate outputs $\mathbf{Y}_{\text{init}}$.
\State Train \texttt{GP} using the dataset $D = \{ \mathbf{X}_{\text{init}}, \mathbf{Y}_{\text{init}} \}$ (including hyperparameter estimation).
\State Let $K = \left\lceil \frac{N - n_0}{B}\right \rceil$ denote the number of acquisition batches.
\For{$k = 1$ to $K$} \Comment{Batched acquisitions}
    \State Initialize a batch of candidate inputs $\mathbf{X}_{\text{cand}} \subset \Xi$ with $|\mathbf{X}_{\text{cand}}| = B$.
    
    \For{iteration = 1 to desired optimization steps} \Comment{Optimize batch EIG}
        \State Compute the Expected Information Gain (EIG) for $\mathbf{X}_{\text{cand}}$ using Eq. (\ref{eq:EIG_closed}).
        \State Update $\mathbf{X}_{\text{cand}}$ to improve the EIG.
    \EndFor
    
    \State Select the optimized batch $\mathbf{X}^*$ with the highest EIG.
    \State Run the simulation $\mathcal{G}$ for each input in $\mathbf{X}^*$ to obtain outputs $\mathbf{Y}^*$.
    \State Add $\{ \mathbf{X}^*, \mathbf{Y}^* \}$ to the dataset $D$.
    \State Refit \texttt{GP} hyperparameters using the updated dataset $D$.
\EndFor

\State Solve the Bayesian calibration problem using MCMC with the GP emulator.
\end{algorithmic}
\end{algorithm}

\begin{algorithm}[H]
\caption{Goal-Oriented Bayesian Optimal Experimental Design (GBOED) with batched acquisitions}
\label{alg:goal_oriented_boed}

\textbf{Input:} Bounds for input space $\Xi$, initial number of samples $n_0$, total simulation budget $N$, batch size $B$, GP emulator \texttt{GP} \\
\textbf{Output:} Samples from the posterior distribution $p(\theta \mid y_{\text{obs}}, \mathcal{GP}(D))$

\begin{algorithmic}[1]
\State Select an initial set of $n_0$ input samples $\mathbf{X}_{\text{init}} \subset \Xi$.
\State Run the simulation $\mathcal{G}$ for each input in $\mathbf{X}_{\text{init}}$ to generate outputs $\mathbf{Y}_{\text{init}}$.
\State Train \texttt{GP} on $D=\{\mathbf{X}_{\text{init}}, \mathbf{Y}_{\text{init}}\}$ (including hyperparameter estimation).

\State Let $K = \left\lceil \frac{N - n_0}{B} \right\rceil$ denote the number of acquisition batches.

\For{$k = 1$ to $K$} \Comment{Batched acquisitions}
    \State Solve the calibration problem using MCMC with the current surrogate model to obtain posterior samples:
    \[
    \mathbf{X}^\prime \sim p(\theta \mid y_{\text{obs}}, \mathcal{GP}(D)).
    \]
    \State Initialize a candidate batch $\mathbf{X}_{\text{cand}} \in \Xi^{B}$ (i.e., $B$ candidate inputs).

    \For{iteration = 1 to desired optimization steps} \Comment{Optimize batch EIG}
        \State Compute the goal-oriented expected information gain for the batch $\mathbf{X}_{\text{cand}}$
        with respect to predictions at $\mathbf{X}^\prime$ via Eq.~(\ref{eq:EIG_closed}).
        \State Update $\mathbf{X}_{\text{cand}}$ to increase $EIG(\mathbf{X}_{\text{cand}})$.
    \EndFor

    \State Select the optimized batch $\mathbf{X}^*=\{\mathbf{x}_1^*,\ldots,\mathbf{x}_B^*\}$.
    \State Run $\mathcal{G}$ at each $\mathbf{x}_b^*$ to obtain outputs $\mathbf{Y}^*=\{y_1^*,\ldots,y_B^*\}$.
    \State Augment the dataset: $D \leftarrow D \cup \{\mathbf{X}^*, \mathbf{Y}^*\}$.
    \State Refit \texttt{GP} on the updated dataset $D$ (including GP hyperparameter re-estimation).
\EndFor

\State Solve the final Bayesian calibration problem using MCMC with the final surrogate:
\[
\mathbf{X}^\prime \sim p(\theta \mid y_{\text{obs}}, \mathcal{GP}(D)).
\]

\end{algorithmic}
\end{algorithm}

\subsection{Calibrate, emulate, sample}
\label{sec:CES}
Calibrate, Emulate, Sample (CES) \cite{cleary2021calibrate} is a sequential framework designed to efficiently tackle Bayesian calibration problems for computationally expensive models. By combining ensemble-based calibration techniques, GP emulators, and MCMC sampling, CES provides a scalable approach to parameter estimation while ensuring robust UQ. It is a practical choice for large-scale or high-dimensional problems and its general framework make it suitable for a variety of applications.

At a high level, CES starts with a calibration phase, in which ensemble Kalman filter techniques such as Ensemble Kalman Inversion (EKI) \cite{kovachki2019ensemble} or Ensemble Kalman Sampling (EKS) \cite{garbuno2020interacting}, are used to obtain an initial  posterior distribution of parameters based on a limited number of model evaluations. The next phase, Emulate, involves training an efficient GP emulator on parameter samples (inputs) and model outputs(s) obtained during the Calibration phase. This emulator is a computationally cheap surrogate for $\mathcal{G}$. 

In the final phase, Sample, MCMC sampling is utilized to further refine the posterior of $\theta$. Employing MCMC in conjunction with the GP emulator enables the drawing of samples from the approximate posterior distribution, thereby ensuring robust UQ. The computational efficiency and inherent smoothness of GP emulators allows for the use of MCMC, which may be prohibitive for expensive $\mathcal{G}$. 

We implement CES following methods specified in \cite{cleary2021calibrate}. In this study we use EKS and EKI as described in \cite{iglesias2013ensemble} for our examples. EKS (unlike most EKI algorithms) does not suffer from collapse of the posterior distribution, but it is computationally expensive and may require more iterations than EKI \cite{garbuno2020interacting}. EKS is typically used only for a limited number of iterations to reduce the computational burden, so the subsequent steps of CES (Emulation and Sampling) are  necessary to obtain reliable UQ and refinement of the posterior distribution. We note that it is also possible to use alternative non-collapsing or regularized variants of EKI. A number of such variants have been developed that help preserve ensemble spread and offer fast convergence. TEKI (Tikhonov-regularized EKI) adds explicit quadratic regularization to the update, mitigating collapse and stabilizing inference in ill-posed settings \cite{chada2020tikhonov}. In addition, Kalman–Bucy/continuous-time EKI with early-stopping rules has been analyzed as a data-driven regularization that halts the dynamics before ensemble collapse \cite{tienstra2024early}. These variants yield final ensembles whose spread better reflects posterior uncertainty and improving computational efficiency relative to running many small EKS steps \cite{tienstra2024early}. Finally, statistically-linearized EKI variants (EKI-SL/IEKF-SL) explicitly prevent collapse in linear tests and maintain empirical covariances near posterior levels \cite{chada2020iterative}. Finally, \cite{huang2022iterated} describe an iterated Kalman framework in which algorithmic choices (discretization, step control) can influence ensemble collapse and computational efficiency.

\subsection{History Matching}
\label{sec:HM}
HM is a statistical methodology that leverages observed data to refine model parameters by ruling out ``implausible'' parameter values based on how well they match the observations \cite{williamson2015identifying}. By iteratively updating parameter configurations, HM effectively narrows the range of plausible values, thereby reducing the computational costs associated with the expensive model evaluations. History matching is well-represented in the field of climate science and has been used extensively for climate model calibration  
\cite{couvreux2021process, williamson2013history, williamson2015identifying,edwards2011precalibrating, raoult2024exploring, lguensat2023semi}.

HM is an iterative process, conducted in ``waves," where regions of the parameter space deemed implausible are progressively eliminated. Implausible regions are parameter configurations that are highly unlikely to produce model outputs resembling the observed data. The remaining parameter space, known as the ``Not-Ruled-Out-Yet" (NROY) space, becomes increasingly refined through successive waves. This iterative narrowing continues until a user-defined stopping criterion is met. 

The first wave is initiated with sampling a limited number of parameter configurations from the entire parameter space. These configurations are used to run the computationally expensive model, and the inputs combined with resulting model outputs form a training dataset to fit a GP emulator. Given the high computational cost of running the model, it is essential to carefully choose a small but representative set of samples. These samples are typically drawn using methods like Optimal Experimental Design (OED) or Latin Hypercube Sampling (LHS) \cite{mckay+bc79}. In our work, we employ LHS because it is straightforward to implement and computationally efficient. For high-dimensional problems, taking advantage of OED for sampling would be better than LHS, because the initial set would be information rich. 

After fitting the GP emulator, we use it to partition the parameter space into plausible and implausible regions, the classification of which is based on an implausibility metric that measures the discrepancy between the model outputs and observed data:

\begin{equation}\label{eq:ImpMetric}
I(\theta) = \frac{||\mathcal{GP}(\theta)- y_{obs}||}{\sqrt{Var(y_{obs})+Var(\mathcal{GP}(\theta))}}.    
\end{equation}

In Eq.(\ref{eq:ImpMetric}), $Var(y_{obs})$ represents the variance of the observations, while $Var(\mathcal{GP}(\theta))$ reflects the uncertainty estimates from the GP model at each input. The metric evaluates how well a parameter configuration reproduces the observed data, accounting for both observational and emulator uncertainties.

We evaluate Eq.(\ref{eq:ImpMetric}) at one million random points in the parameter space, which ensures that enough points remain for subsequent waves after each wave of history matching. Regions of the parameter space where the implausibility metric is above a certain threshold are discarded, as those parameter configurations are unlikely to have generated the observed data. A threshold of three suffices due to the ``three-sigma'' rule defined in \cite{pukelsheim94}, which states that for any continuous, unimodal distribution, $95\%$ of its probability will lie within $\pm 3\sigma$ of the mode regardless of the distribution's asymmetry, skewness, or heavy tails (see Algorithm 2.3).

Subsequent waves refine the NROY space further. For each wave, new training samples are drawn from within the current NROY space, the expensive model is run, and the emulator is updated. Sampling from the NROY space is challenging because uniform sampling is infeasible when the space becomes fragmented due to eliminated ``clusters of inputs''. Instead, we sample uniformly from the remaining points already identified as part of the NROY space. The HM process continues until a stopping criterion is met. This could be based on computational constraints or when the NROY space becomes too small to warrant further refinement. In our implementation, we use a fixed threshold of three for the implausibility metric in all waves and rely on the available computational budget to determine when to stop.  To automate the HM process, \cite{drovandi+np21} proposes a semi-automated approach using sequential MCMC, which eliminates the need for user-specified parameters like implausibility thresholds.  Instead, we adhere to a simpler framework with fixed thresholds and computational constraints guiding our stopping criteria.

The HM algorithm is performed first to generate a set of informative input–output pairs. After the algorithm terminates, all inputs and corresponding model outputs generated across all waves are collected, and Gaussian process emulators are fitted to these data, as described in Section 2.1. The final emulators are then used within an MCMC framework to construct the posterior distribution over calibration parameters. Within this final MCMC step, both the GP predictive mean and predictive covariance are used to define the likelihood (Eq. \ref{eq:likelihoodwGP}). 

\begin{algorithm}
\caption{History Matching Algorithm}
\label{hmalgo}
\begin{algorithmic}[1]
\State \textbf{Input:} Initial set of parameter sample bounds $bounds$, expensive model $\mathcal{G}$, observations $y_{obs}$
\State \textbf{Output:} Refined Not ruled out yet (NROY) space
\State Generate uniform samples of the parameters within the bounds, $\theta_{bounds}$.
\State Generate corresponding outputs $\mathcal{G}(\theta_{bounds})$ for $\theta_{bounds}$.
\State Train GP emulator using samples from the simulator, $(\theta_{bounds}, \mathcal{G}(\theta_{bounds}))$
\For{each wave of history matching}
\For{each column of outputs}
    \State Compute implausibility metric $I(\theta)$ for each sample using Eq. (\ref{eq:ImpMetric}).
    \State Remove regions of parameter space with $I(\theta) > 3$
    \EndFor
    \State Take the intersection of the NROY space across all outputs
    \State Sample new points uniformly from the refined NROY space
    \State Retrain the GP emulator with the new samples
\EndFor
\State \textbf{End} when stopping criterion (e.g., computational budget) is met
\end{algorithmic}
\end{algorithm}

\section{Lorenz'96 Multiscale system}
\label{sec:Lorenz}
We first use the Lorenz '96 multiscale system described in \cite{lorenz1996predictability} (Eq. \ref{eqn:lorenz96}) to test calibration methods. The Lorenz '96 multiscale system is often used to study the behavior of chaotic systems such as climate \cite{lguensat2023semi, ott2004local, orrell2003model, parthipan2023using}. The system consists of a set of ordinary differential equations that describe the evolution of a large number of interacting variables. In this system, there are $K$ large-scale variables ($X_k$) and  $JK$ small-scale variables ($Y_{j,k}$). The $X$ and $Y$ variables interact with each other, but there is also coupling between them, as shown in Eq. (\ref{eqn:lorenz96}).

\begin{equation} \label{eqn:lorenz96}
    \begin{aligned}
        \frac{{dX_k}}{{dt}} = -X_{k-1}(X_{k-2} - X_{k+1}) - X_k - \frac{hc}{b} \ \sum_{j=1}^J Y_{j,k} + F \\
        \frac{dY_{j,k}}{dt} = -cb Y_{j+1, k} (Y_{j+1,k} - Y_{j-1, k}) -cY_{j,k} +\frac{hc}{b}  X_k.  
    \end{aligned}
\end{equation}

The large- and small-scale variables are periodic over $J$ and $K$, i.e., $X_{k-K}, X_{k+K} = X_k$, and $Y_{j, k-K}, Y_{j, k+K} = Y_{j_K}$ and $Y_{j-J,k} =Y_{j, k-1} $, and $Y_{j+J,k} = Y_{j,k+1}.$ The parameters of the system, $\theta = (h, F, c, b)$, control the degree of forcing, coupling, and amplitude of the system. 

We perform Bayesian inversion for $\theta$ based on data averaged across the $K$ locations and over time windows of length $T$ = 100 for all methods. We consider the following observables arising from the $k-$indexed operator $\varphi_k: \mathbb{R} \times \mathbb{R}^J \rightarrow \mathbb{R}^5$ as defined in \cite{cleary2021calibrate}: $\phi_k(Z) = \phi(X_k, Y_{1,k}, \dots, Y_{J, k}) = (X_k, \Bar{Y}_k, X_k^2, X_k \Bar{Y}_k, \Bar{Y}_k^2)$, where $Z$ characterizes the current state of the system. 

Generally, it can be assumed that the observations arising from this system can be expressed as a forward operator $\mathcal{G}(\cdot)$ acting on input parameters $\theta$ plus noise:
\begin{equation}\label{eq:ModelForm}
    y=\mathcal{G}(\theta) + \eta,
\end{equation}
where $\eta \sim N(0, \Gamma_y(\theta))$ and $\Gamma_y(\theta) = T^{-1} \Sigma(\theta)$. Typically, $\Gamma_y(\theta)$, which captures variability due to initial conditions, must be approximated by $\Gamma_{obs}$. To obtain $\Gamma_{obs}$ we simulate the Lorenz '96 system for a $\tau=40,000$ time steps and calculate the covariance over windows of $T=100$, in accordance with \cite{cleary2021calibrate}. 
The forward operator is then defined as:
\begin{equation*}
    \mathcal{G}(\theta) = \frac{1}{T} \int_0^T \left ( \frac{1}{K} \sum_{k=1}^K \phi_k(Z(s) ) \right ) ds.
\end{equation*}

\subsection{Performance metrics}
\label{sec:Validation}
We consider two validation metrics, absolute error (AE) and proper (log) score (Eq. 25 in \cite{gneiting2007strictly}), which balances accuracy in point estimates as well as UQ. In the equations for AE and score below, $\hat{\mu}$ and $\hat{\sigma}$ are the estimated means and standard deviations for components (e.g. $h,F,c,b$) of $\theta$, and $y$ represents the corresponding true value.  Score and AE are calculated as follows:
\begin{align*}
\mathrm{AE}(y,\hat{\mu}) &= |y - \hat{\mu}|, \\
\mathrm{score}(y, \hat{\mu}, \hat{\sigma}) &= 
- \log(\hat{\sigma}^2)  - (y - \hat{\mu})^\top \frac{1}{\hat{\sigma}^{2}} (y - \hat{\mu}). 
\end{align*}

Smaller AEs and larger scores are preferred. Prior to calculating validation metrics, values for the parameter $\mathrm{log}c$ of the Lorenz '96 system were exponentiated. Because all methods provide a posterior distribution for components of $\theta$, we use the empirical means and standard deviations as point estimates for each component of $\theta$ in computing validation metrics. 

\subsection{Implementation details}
\label{sec:Experiment}
To assess how the performance of methods changed with increasing model evaluations, we performed calibration twice with each method, using relatively few (200) and more (1000-5400) model evaluations. We chose 200 model evaluations based on methodology in  \cite{yarger2024autocalibration}, and 1000 represents a reasonable upper bound for the number of model evaluations that would be practically feasible with an expensive climate model \cite{dunbar2021calibration}. To obtain the final posterior distribution for $\theta$ for all methods, we use GP emulators, along with corresponding $\phi_k(Z)$ and $\Gamma_{obs}$ to perform MCMC sampling. We use the same prior specification for GP lengthscales as specified in \cite{cleary2021calibrate} for all methods for consistency.  MCMC sampling was conducted via Sequential Tempered Markov Chain Monte Carlo (ST-MCMC) \cite{catanach2020bayesian}, which is a highly effective Sequential Monte Carlo (SMC) sampler due to its efficient parallelism while retaining sampling robustness due to its adaptivity and monitoring of sample population statistics instead of individual chain statistics. We found that 8092 ST-MCMC parallel samples resulted in a comprehensive exploration of the posterior for all methods. See \cite{catanach2020bayesian} for more details on ST-MCMC sampling. 

\subsubsection{CES and EKS}
For both examples, we use EKS with an adaptive timestep as described in \cite{garbuno2020interacting}, which is also the version implemented in the original CES paper \cite{cleary2021calibrate}. EKS was implemented using the Julia package EnsembleKalmanProcesses.jl \cite{climateAlliance} using default settings. We did not impose strict bounds on the prior in our implementation of EKS. For performing Bayesian inversion on the Lorenz '96 system (Eq. \ref{eqn:lorenz96}) via CES, EKS was run for 54 iterations with 100 ensemble members (5400 model evaluations), in accordance with \cite{cleary2021calibrate}. We also ran EKS with only 10 iterations and 20 ensemble members (200 model evaluations). For the case of 5400 model evaluations, only iterations $1, 6, 12, \dots, 54$ were saved for subsequent GP fitting, also as suggested in \cite{cleary2021calibrate}. Doing so reduces the total number of data required to fit GPs, decreasing computation time while retaining the spread of the samples over the total number of EKS iterations. 

\subsubsection{BOED and GBOED}
For both BOED and GBOED with 200 or 1000 total model evaluations, the Gaussian process (GP) surrogate was constructed in a batched sequential design framework. An initial GP model was trained using 20 prior Monte Carlo samples for coverage of the input domain. Thereafter, additional design points were selected in batches of 20, with the GP surrogate retrained after each batch. Within each iteration, candidate design points were optimized using L-BFGS-B with gradients computed using JAX automatic differentiation. These steps were repeated until the target evaluation budget was reached as described in Algorithms \ref{alg:standard_boed} and \ref{alg:goal_oriented_boed}. For GBOED, the intermediate sampling required to approximate the expected information gain was performed using ST-MCMC \cite{catanach2020bayesian}.

\subsubsection{HM}
For HM, we used both 40 model evaluations across five waves (200 total evaluations) and 200 model evaluations across 5 waves (1000 total evaluations). To obtain the final posterior distribution of the parameters, we use the final trained GP emulators to perform ST-MCMC sampling. 

For reporting results, all methods are named using the method name (i.e., BOED, GBOED, CES, HM) followed by a suffix indicating the number of model evaluations. For example, CES5400 refers to the CES method where $\mathcal{G(\cdot)}$ is evaluated 5400 times.  

\subsection{Experiments}
In this section, we describe experiments pertaining to calibration of the Lorenz '96 multiscale system using the main methods CES, HM, BOED, and GBOED. First, we examine the marginal posterior distributions of $h, F$, $c$ and $b$. We then conduct a convergence analysis of all methods over increasing model evaluations, and finally, we conduct an in-depth performance assessment comparing accuracy and UQ of all methods. 

\begin{figure}[H]
    \centering
    \includegraphics[scale=.46]{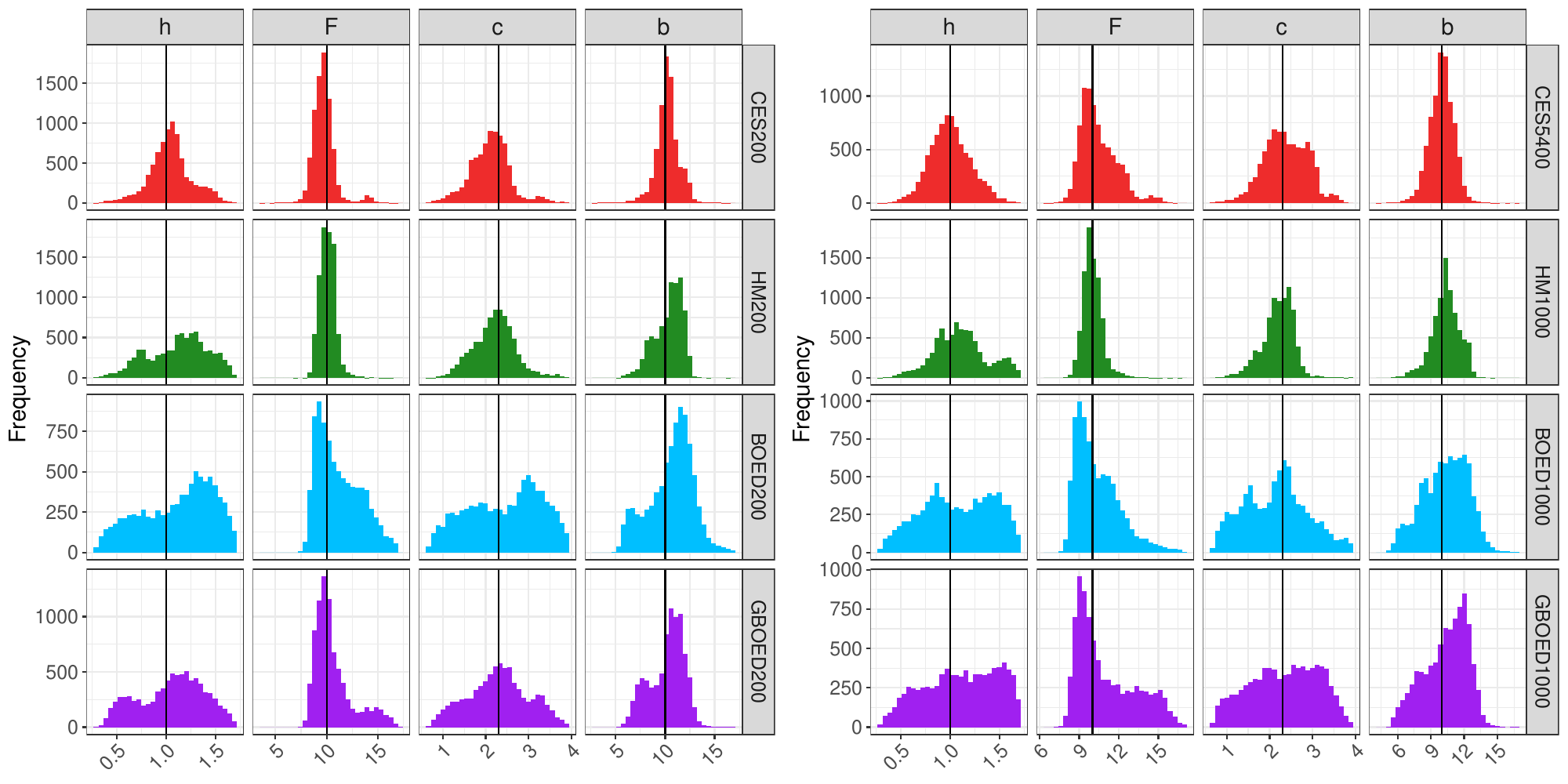}
    \caption{Frequency histograms of posterior distributions of parameters $h, F, \log c$, and $b$ resulting from CES200, HM200, BOED200, and GBOED200 (left) and CES5400, HM1000, BOE1000, and GBOED1000 (right). Histograms represent 8092 values. True parameter values are denoted by vertical lines.}
    \label{fig:cespost200}
\end{figure}

First we examine the posterior distribution resulting from all methods for $\theta = (1,10, 10, 10)$. Posterior distributions for $h, F, \textup{log}c$ and $b$ for the case of fewer (200) and more (1000-5400) model evaluations are shown in the left and right panels, respectively, of Figure \ref{fig:cespost200} (recall that CES was run with 5400 model evaluations following methods in \cite{cleary2021calibrate}). Due to the nonlinear and chaotic nature of the Lorenz '96 system and correlation among parameters, the resulting posterior distributions are not expected to be Gaussian. 

For this particular configuration of the Lorenz '96 system, the posterior for CES and HM are tightest around the true values, and the shape of the posterior remains very similar for either 200 (\ref{fig:cespost200}, left panel) or 1000-5400 (\ref{fig:cespost200}, right panel) model evaluations. For all methods, posterior distributions for parameter $b$ tend toward Gaussian, but posteriors for $F$ are more right-skewed. Despite differences in posteriors, all methods were able to capture the true parameter values (denoted by black vertical lines in Figure \ref{fig:cespost200}). Note also that as the number of model evaluations increases, it may not necessarily be the case that the variance of the posterior distribution will decrease. As model evaluations increase, methods such as CES and GBOED especially may explore previously under-sampled regions of parameter space, leading to a more realistic representation of uncertainty. 

\paragraph{Convergence analysis}
\label{sec:convergence_analysis}
 Next we conducted an exercise to gain insight into convergence rates of the main methods in calibrating the Lorenz '96 system for the same configuration ($\theta = (1,10,10,10)$). The goal of the convergence exercise was to determine how performance metrics of each method changed over increasing model evaluations. We report both score and AE every 100 evaluations up to 1000 total model evaluations for all methods. We included the intermediate (calibration) step of CES (EKS) as well for comparison.

\begin{figure}[H]
    \centering
    \includegraphics[scale=0.8]{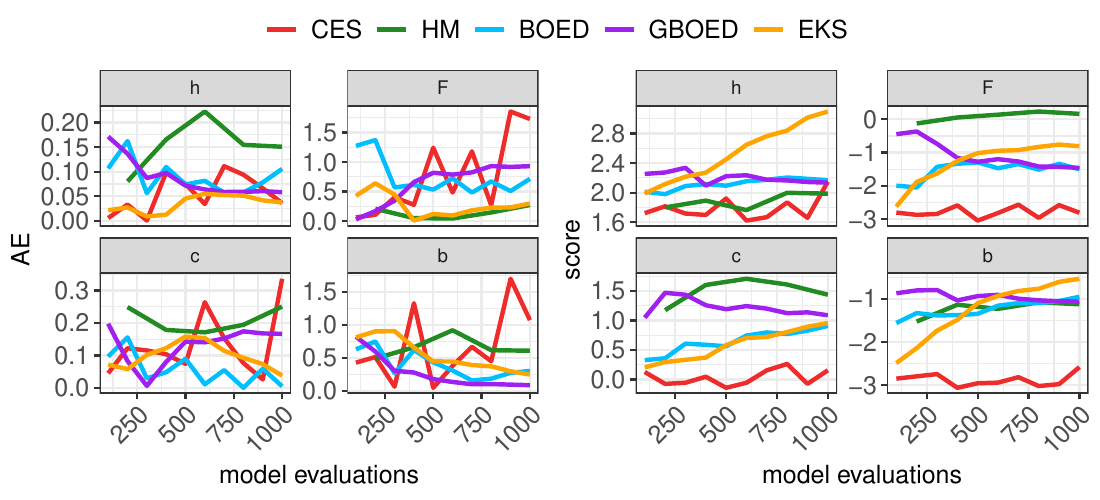}
    \caption{AEs (left) and scores (right) for CES, EKS, HM, BOED, and GBOED over increasing model evaluations for calibrating the Lorenz '96 system using base parameters $\theta = (1, 10, 10, 10)$ (results for parameter $c$ are calculated in log space).}
    \label{fig:convergenceCESEKS}
\end{figure}

AEs generally decreased as the number of model evaluations increased for parameter $b$, but there was considerable variability among methods for parameters $h$, $F$ and $\log c$ (Figure \ref{fig:convergenceCESEKS}, left panel). Scores generally tended to improve with increasing model evaluations, but this was not consistent for all methods and parameters. EKS, BOED, and to some extent HM, were the only methods that demonstrated improvement in scores over increasing model evaluations (Figure \ref{fig:convergenceCESEKS}, right panel). Scores for CES were relatively stable, while those for GBOED decreased very slightly from 100 to 1000 model evaluations. 

While one might expect performance to improve monotonically with increasing model evaluations, convergence trends varied across parameters and methods, with performance often plateauing after an initial improvement. This behavior may reflect a combination of surrogate saturation, parameter non-identifiability, and irreducible variability associated with finite-time statistics in the chaotic Lorenz ’96 system. Notably, only EKS exhibited consistent improvement in score with increasing evaluations, which is expected given its ensemble-based contraction mechanism that directly reduces posterior variance. In contrast, methods relying on emulator-based design or filtering exhibited diminishing returns once calibration-relevant regions were adequately explored. This convergence analysis suggests that CES, GBOED/BOED, and HM may require fewer model evaluations for chaotic systems such as the Lorenz '96 multiscale system. Finally, it is important to note that these conclusions are based on a \emph{single} set of true parameters for this particular system. 

\paragraph{Performance evaluation}
To obtain a fair and rigorous comparison of the main methods, we performed Bayesian inversion for $\theta_i, i = 1\dots 60$ using the Lorenz '96 multiscale system (Eq. \ref{eqn:lorenz96}). Each component of $\theta_i$ was drawn from the following prior distributions based on baseline parameter values of 1, 10, $\textup{log}$10, and 10 for $h, F$, $\log c$, and $b$, respectively:
\begin{align*}
    h \sim \mathrm{Unif}(0.3, 7) \\
    F \sim \mathrm{Unif}(3, 17) \\
    \log c \sim \mathrm{Unif}(0.6908,  3.9144)\\
    b \sim \mathrm{Unif}(3, 17).
    \label{eq:prior}
\end{align*}
Note that parameter $c$ must be non-negative, so inversion for this parameter is performed in log space. Each component of $\theta_i$ was drawn from a Uniform distribution with the minimum (maximum) being the baseline parameter value minus (plus) 70\% of the baseline value. While these priors are more constrained than the Gaussian priors used in \cite{cleary2021calibrate}, we found that Uniform prior distributions ensured stability of the system during calibration. Observables from this system  ($\phi_k(Z)$) and their associated variability ($\Gamma_{obs}$)  were computed for each $\theta_i$ and were used within the MCMC sampling step included within all methods to obtain the posterior distribution for each $\theta_i$.
We also compared the four main methods BOED, GBOED, CES, and HM against the following competitors:
\begin{itemize}
\item \underline{LHS200/2000:} Posterior parameter distributions obtained via parameter values generated using a LHS design combined with MCMC. First, 200 (LHS200) or 1000 (LHS1000) LHS samples for parameters were generated. Next, the Lorenz '96 system was run using the LHS samples, and observed quantities $\phi_K(Z)$ were computed for each of the 200 or 1000 samples. LHS parameter samples (inputs) and the observables (outputs) were used for GP emulator fitting. Finally, the fitted GPs, along with observables $\phi_k(Z)$ and $\Gamma_{obs}$ corresponding to each $\theta_i$ were used to perform ST-MCMC sampling \cite{catanach2020bayesian}. Including LHS-based methods as comparators allows us to determine if the calibration-specific steps of each method (e.g. EKS in CES) have added benefits. 
\item \underline{EKS200/5400:} the posterior distribution resulting from EKS (the Calibrate step of CES). 
We include EKS, because, in contrast to other ensemble Kalman Filter methods such as EKI, it can provide systematic UQ during calibration \cite{cleary2021calibrate}. We use the final iteration of EKS as the posterior distribution for calculating validation metrics. No MCMC sampling (and therefore no GP emulation) is required for EKS.
\end{itemize}

\paragraph{Statistical analyses}
\label{sec:AssessPerformance}
To evaluate how well different methods performed, we constructed two linear, additive models: one for AE and another for score. The response for the models was either AE or score, and the two covariates were the type of method (e.g., BOED200, CES54000, etc) ($METHOD$) and parameters of the Lorenz '96 model ($h, F, c, b$) ($PARAM$). Thus, $METHOD$ and $PARAM$ are factor-valued covariates with 11 and 4 levels, respectively. We considered AE and scores from the 60 calibration problems of the Lorenz '96 system as replicates in these models. The linear models for AE and score are thus two-way, additive analysis of variance (ANOVA) models with no interaction terms. We did not include interaction terms to preserve degrees of freedom, and preliminary analyses showed that interactions were not practically significant. The general form for both models can be expressed as:

\begin{equation*}
    y_{ijk} =\mu + \alpha_j + \beta_k + \epsilon_{ijk},
\end{equation*}
where $y_{ijk}$ is the observation for the $ith$ replicate , $jth$ level of $METHOD$ and the $kth$ level of $PARAM$; $\mu$ is the overall mean, $\alpha_j$ and $\beta_k$ are the effects of the $jth$ level of $METHOD$ and $kth$ level of $PARAM$, respectively, and $\epsilon_{ijk}$ is a normally distributed error term. The two ANOVA models for AE and score can be written more specifically as:
\begin{align}\label{eq:LinearModels}
    y_{ijk}^{AE} = \beta_0 + \beta_1^{(METHOD_j)} + \beta_2^{(PARAM_k)} + \epsilon_{ijk},\\
    y_{ijk}^{SCORE} = \beta_0 + \beta_1^{(METHOD_j)} + \beta_2^{(PARAM_k)} + \epsilon_{ijk},\nonumber
\end{align}
where, in this formulation, $\beta_1$ represents the effects of the 10 levels of $METHOD$ relative to the $reference$ level, and $\beta_2$ represents the effects of the three levels of $PARAM$ relative to the reference level. This model can be generally expressed in matrix form as $\mathbf{Y} = \mathbf{X}\boldsymbol{\beta} + \boldsymbol{\epsilon}$, where $\mathbf{X}$ is a design matrix that encodes the levels of $METHOD$ and $PARAM$. The coefficient vector $\boldsymbol{\beta}$ contains ten coefficients for $METHOD$, three coefficients for $PARAM$, and one for the intercept.  Additionally, we excluded extremely low score values (less than -50, primarily from LHS methods) from the dataset, as these could have disproportionately influenced the results.  In total, 10 observations were removed. After fitting these models, we performed post-hoc pairwise comparisons of the estimated marginal means of AE by $METHOD$ to determine if differences between any two calibration methods were significantly different. Pairwise comparisons between methods were performed using Tukey’s honestly significant difference (HSD) procedure. Significance of pairwise tests was assessed at $\alpha = 0.05$. All models were fitted in R using the \texttt{lm()} function \cite{baseR}; pairwise comparisons were carried out using the \texttt{emmeans} \cite{emmeansPackage} and \texttt{modelbased} \cite{modelbasedPackage} packages.

\begin{figure}[H]
    \centering
    \includegraphics[scale=.4]{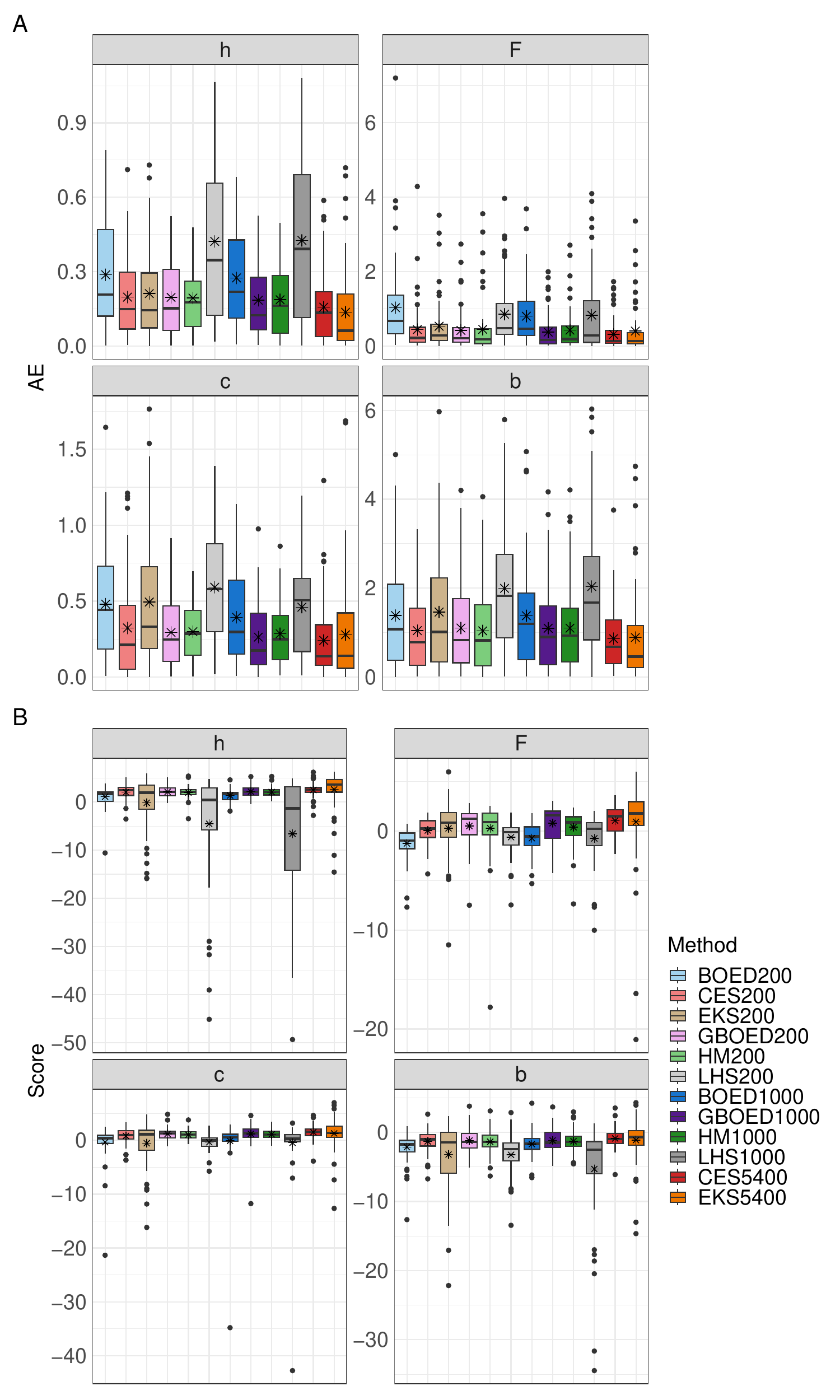}
    \caption{ Boxplots of AEs (A) and log score (B) for calibrated parameters $h, F, \log c$ and $b$ of the Lorenz '96 model for all methods. Each boxplot for AE represents 60 values; values less than -50 for score were omitted. AE values were calculated using the mean of the posterior distribution. Horizontal lines within boxplots represent the median, while stars denote means.}
    \label{fig:rmsescoreBoxplots}
\end{figure}

\paragraph{Results}
\label{sec:results}
All methods except BOED- and LHS-based approaches yielded reasonably accurate point estimates of all parameters (Figure \ref{fig:rmsescoreBoxplots}A) and provided a good balance between accuracy and uncertainty quantification (Figure \ref{fig:rmsescoreBoxplots}B). LHS-based methods, in particular, consistently underperformed in terms of both accuracy and UQ, emphasizing the limitations of relying solely on prior-based sampling without incorporating a data-informed calibration step (Figure \ref{fig:rmsescoreBoxplots} A and B).

\begin{figure}[ht]
    \centering
    \includegraphics[scale=.6]{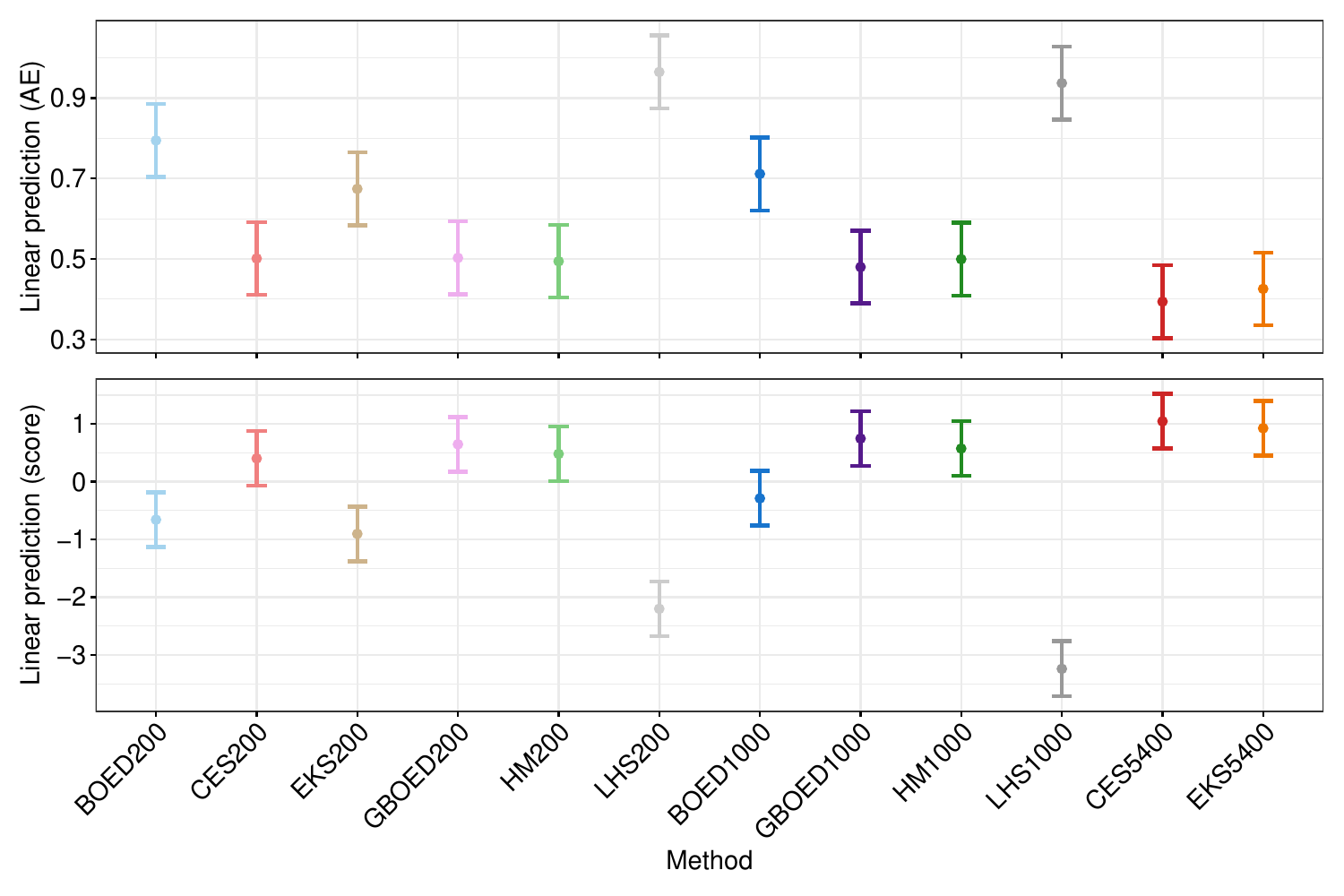}
    \caption{Estimated marginal means of absolute errors (AEs) (top) and score (bottom) averaged over the four parameters for the \emph{AE} and \emph{score} linear models described in Section 3.2. Error bars represent 95\% confidence intervals (note that the confidence intervals reflect uncertainty around the estimated marginal means, not the raw data).}
    \label{fig:rmsemodel}
\end{figure}

Pairwise comparisons of estimated marginal means of AE (averaged over all parameters) from the linear AE model (Eq. \ref{eq:LinearModels}) revealed notable performance trends. LHS-based methods (LHS200, LHS1000) and BOED200 performed significantly worse than all other methods (Table \ref{tab:betterthanRMSEScore}, Figure \ref{fig:rmsemodel}). CES5400, EKS5400, and GBOED1000 formed a top-performing group, each significantly outperforming BOED200, BOED1000, EKS200, LHS200, and LHS1000 (Figure \ref{fig:rmsemodel}, Table \ref{tab:betterthanRMSEScore}). HM methods also performed reasonably well; both HM200 and HM1000 performed similarly and achieved significantly lower AEs than BOED and LHS methods. 

The effect of increasing model evaluations varied by method. For EKS, mean AE decreased by approximately 50\% from 200 to 5400 evaluations, and this difference was statistically significant (Table SM1, Figure \ref{fig:rmsemodel}, top panel). For BOED and CES, mean AEs decreased with increasing evaluation count (from 0.85 to 0.72 for BOED and from 0.50 to 0.29 for CES), although these reductions were not statistically significant. Increasing model evaluations for HM and GBOED resulted in marginal decreases in AEs (Figure \ref{fig:rmsemodel}, top panel). 

Out of the methods utilizing fewer evaluations (200), GBOED200, CES200, and HM200 performed best. Mean AEs of these methods were very similar, and although they were larger than those of the best-performing methods (e.g., CES5400, EKS5400, GBOED1000), the differences were not statistically significant. Among the methods with higher evaluation counts, CES5400 achieved the lowest mean AE (Figure \ref{fig:rmsemodel}, top panel). Another notable finding was that for fewer (200) model evaluations, CES resulted in considerable improvement over EKS. However, for more model evaluations, improvements were marginal (Figure \ref{fig:rmsemodel}, top panel). All pairwise comparisons from the linear AE model are provided in Table SM1.

Results for score were largely consistent with those for AE, with some notable differences. Like results for AE, LHS200 and LHS1000 had significantly worse (lower) scores compared to all other methods (Table \ref{tab:betterthanRMSEScore}, Figure \ref{fig:rmsemodel}, bottom panel). LHS-based methods were the only approaches exhibiting large, negative outliers (-50 or less). This finding emphasizes the importance of an initial calibration step to achieve robust UQ. CES5400 again achieved the highest mean score, significantly outperforming five methods: LHS200, LHS1000, BOED200, BOED1000, and EKS200. However, EKS5400, GBOED200, and GBOED1000 also performed very well and outperformed the same five methods as CES5400 (Table \ref{tab:betterthanRMSEScore}, Figure \ref{fig:rmsemodel}). HM200 and HM1000 had nearly identical mean scores and performed only slightly worse than top performing methods. All pairwise comparisons from the linear score model are provided in Table SM2. As an additional consistency check, we also evaluated the joint posterior density at the true parameter values for each of the replicates using a multivariate kernel density estimate; results are broadly consistent with those obtained using AE and score (see SM7).

\begin{table}[H]
\footnotesize
\caption{Methods in the \emph{leftmost} column have significantly lower (better) mean AE values than those in the middle column (AE) and significantly higher (better) scores than those in the rightmost column (Score) at $\alpha = 0.05$. For example, the estimated marginal mean AE of EKS200 in the leftmost column is significantly lower than mean AEs LHS200 and LHS1000 in the middle column. Significant differences are based on pairwise comparisons from the linear models for AE and score (Eq. \ref{eq:LinearModels}).}
\label{tab:betterthanRMSEScore}
\begin{tabular}{|l|l|l|}
\hline
\multicolumn{1}{|c|}{\textbf{Method}} & \multicolumn{1}{c|}{\textbf{AE}} & \multicolumn{1}{c|}{\textbf{Score}} \\ \hline
BOED200 & none & LHS200, LHS1000 \\ \hline
BOED1000 & LHS200 & LHS200, LHS1000 \\ \hline
CES5400 & \begin{tabular}[c]{@{}l@{}}BOED1000, BOED200, \\ EKS200, LHS200, LHS1000\end{tabular} & \begin{tabular}[c]{@{}l@{}}BOED1000, BOED200  \\ EKS200, LHS200, LHS1000\end{tabular} \\ \hline
CES200 & LHS200, LHS1000 & LHS200, LHS1000 \\ \hline
EKS5400 & \begin{tabular}[c]{@{}l@{}}BOED1000, BOED200,\\  EKS200, LHS200,\\  LHS1000\end{tabular} & \begin{tabular}[c]{@{}l@{}}BOED1000, BOED200,\\ EKS200, LHS200,\\  LHS1000\end{tabular} \\ \hline
EKS200 & \begin{tabular}[c]{@{}l@{}} LHS200, LHS1000\end{tabular} & \begin{tabular}[c]{@{}l@{}}LHS200,\\ LHS1000\end{tabular} \\ \hline
GBOED1000 & \begin{tabular}[c]{@{}l@{}}BOED1000, BOED200,\\ EKS200, LHS200, LHS1000\end{tabular} & \begin{tabular}[c]{@{}l@{}}BOED1000, BOED200,\\ LHS200, LHS1000, EKS200\end{tabular} \\ \hline
GBOED200 & \begin{tabular}[c]{@{}l@{}}BOED200, \\ LHS200, LHS1000 \end{tabular} & \begin{tabular}[c]{@{}l@{}}BOED200, BOED1000,\\EKS200, LHS200, LHS1000\end{tabular} \\ \hline
HM200 & \begin{tabular}[c]{@{}l@{}}BOED200, BOED1000,\\ LHS200, LHS1000\end{tabular} & LHS200, LHS1000, EKS200, BOED200 \\ \hline
HM1000 & \begin{tabular}[c]{@{}l@{}}BOED200, BOED1000,\\ LHS200, LHS1000\end{tabular} & LHS200, LHS1000, BOED200, EKS200 \\ \hline
LHS200 & none & LHS1000 \\ \hline
LHS1000 & none & none \\ \hline
\end{tabular}
\end{table}

\section{Two-layer quasi-geostrophic (QG) system}
\label{sec:QGsystem}
For our second example, we consider the quasi-geostrophic system (QG) \cite{fandry1984two}. It provides a good test for calibration methods, because this system is computationally demanding \cite{springer2021efficient}. We use the two-layer QG model implemented in \texttt{GeophysicalFlows.jl} \cite{constantinou2021geophysicalflows}. The prognostic variables are the potential vorticities (PVs) $q_1(x,y,t)$ and $q_2(x,y,t)$ in an upper and lower layer of thicknesses $H_1$ and $H_2$, respectively. Standard QG processes are included: nonlinear advection, planetary vorticity gradient ($\beta$-effect), interlayer coupling set by stratification, linear bottom drag $\mu$ in the lower layer, and small-scale dissipation via hyperviscosity $\nu$. A mean zonal shear $U$ is imposed to trigger baroclinic instability; after spin-up the flow reaches a statistically stationary eddy field.

We solve on a doubly periodic domain of size $L_x \times L_y = 2\pi \times 2\pi$ using a pseudo-spectral method with $N_x \times N_y = 256 \times 256$ Fourier modes, 2/3 de-aliasing, and time step $\Delta t=1e^{-3}$. The model is integrated in time with a fourth-order exponential time–differencing Runge–Kutta scheme. Unless stated otherwise we non-dimensionalize by the domain scale and planetary vorticity, and use double precision. All QG runs in this paper use the same discretization and boundary conditions. The QG system has been used to test Bayesian calibration methods as in \cite{springer2021efficient}, but in contrast to the study in \cite{springer2021efficient}, we calibrate four parameters ($H_1, H_2, \mu, \nu$) rather than only $H_1$ and $H_2$.

\subsection{Implementation details}
We model observations as $y_{obs}=\mathcal{G}(\theta)+\varepsilon$, $\varepsilon\sim\mathcal{N}(0,\Gamma_{obs})$, where $\mathcal{G}(\theta)\in\mathbb{R}^{10}$ denotes the vector of ten flow statistics derived from the QG model at parameter value $\theta$ (for exact definitions see SM8). To compute $y_{obs}$, we first integrate for a spin-up of 75,000 steps to statistical stationarity. Then, the ten flow statistics are calculated over windows of length 25,000 step until 200 samples are collected. The observation vector, $y_{obs}$, is set to be the final (200th) sample. The observed covariance matrix, $\Gamma_{obs}$ is computed as the covariance over the 200 samples. This procedure to obtain $y_{obs}$ and $\Gamma_{obs}$ is thus similar to the procedure used to calculate these quantities for the Lorenz '96 example. 
We calibrate
\[
\theta = (\mu, \nu, H_1, H_2),
\]
with $\log$-parameters for $(\mu,\nu)$ to enforce positivity. $H_1$ and $H_2$ control the stratification and the baroclinic coupling strength (i.e., the deformation radius). They influence how energy and PV variance partition between layers and the preferred mesoscale length scales. The parameter $\mu$ is a large-scale linear damping acting primarily in the lower layer. It sets a natural damping timescale, and $\nu$ suppresses small-scale variance and extreme PV excursions. Together, $H_1, H_2, \mu,$ and $\nu$ shape the energy budget, scale selection, interlayer coupling, and intermittency, making them natural targets for Bayesian calibration. The $\beta$ parameter and imposed shear $U$ are fixed at $\beta=5$ and $U=[1,0]$. We consider the true parameters $\theta = (0.032, 9.5 \times 10^{-6}, 0.25, 0.85)$ for $\mu, \nu, H_1$ and $H_2$. Priors are independent uniform distributions: 
\begin{align*}
    \log\nu\sim \textup{Unif}(\log(5\times 10^{-6}), \log(1\times 10^{-4})) \\
    \log\mu\sim \textup{Unif}(\textup{log}(0.01), \textup{log}(0.2)) \\
    H_1\sim \textup{Unif}(0.1, 0.3) \\
    H_2\sim \textup{Unif}(0.6, 0.9).
    \label{eq:prior}
\end{align*}

Because the QG system is computationally demanding, we calibrate it for a single parameter setting rather than across multiple replicates, as was done for the Lorenz ’96 example. We nevertheless examine the effect of computational budget by comparing results obtained using 200 and 500 QG model evaluations.

For QG calibration, we use Ensemble Kalman Inversion (EKI) \cite{iglesias2013ensemble, schillings2017analysis} rather than EKS. In preliminary tests, EKI produced more stable and accurate parameter estimates than EKS for moderate ensemble sizes and computational budgets. EKI has also been used successfully as the calibration step in CES for estimating parameters in general circulation models \cite{dunbar2021calibration, dunbar2022ensemble}. The improved performance of EKI relative to EKS in this setting may be attributable to the additional stochasticity introduced by EKS combined with mild confounding between parameters $H_2$ and $\nu$, which can lead to biased ensemble drift. We implement EKI using the Julia package \emph{EnsembleKalmanProcesses.jl} \cite{climateAlliance} with default settings.

For the 200-simulation case, we use $N=40$ ensemble members and 5 EKI iterations, while for the 500-simulation case we use $N=100$ ensemble members and 5 iterations, which aligns with other studies that use EKI for the calibration step of CES \cite{howland+ds22, dunbar2021calibration, dunbar2022ensemble}. Statistics for ensemble members are computed in parallel using GPUs. After completion of the EKI iterations, GP emulators are trained using all ensemble inputs and corresponding outputs (the 10-dimensional vector of QG statistics), and posterior sampling is performed using ST-MCMC \cite{catanach2020bayesian}. This MCMC procedure is applied consistently across CES and all other methods.

History Matching (HM) is performed in the same manner as for the Lorenz ’96 system, using five waves with 40 evaluations per wave for the 200-evaluation case and 100 evaluations per wave for the 500-evaluation case. BOED and GBOED are implemented analogously, using 10 and 25 batches of 20 QG evaluations, respectively, yielding total budgets of 200 and 500 evaluations. The batch size is chosen to match the HPC job limit, allowing all model evaluations within a batch to run concurrently on separate nodes. For all methods, the final posterior is obtained by training GP emulators on the full set of input–output pairs generated during calibration and applying ST-MCMC, as in the Lorenz ’96 example.

\subsection{Results}
\begin{figure}[H]
    \centering
    \includegraphics[scale=.42]{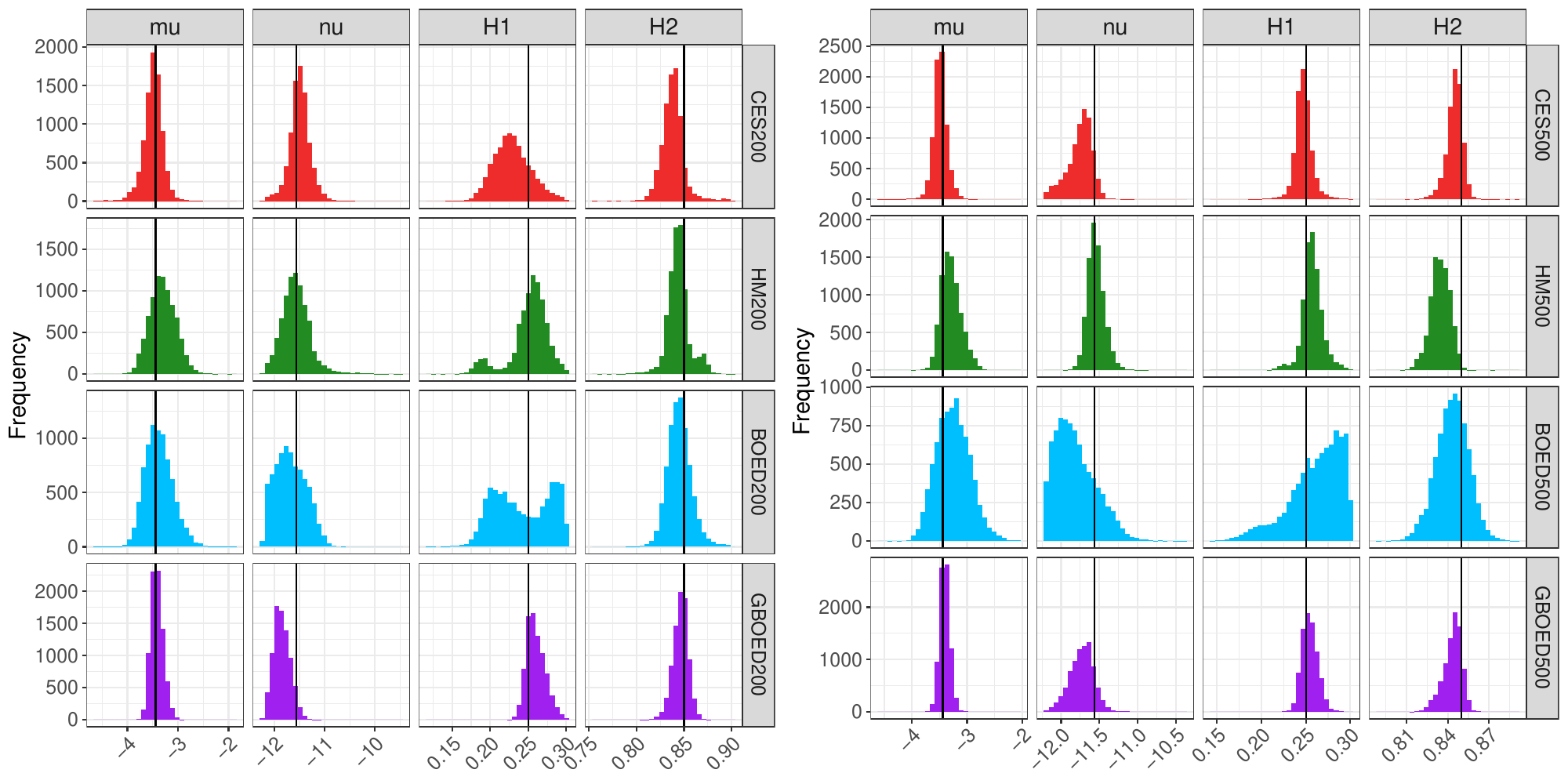}
    \caption{Frequency histograms of posterior distributions of $\textup{log}\mu, \textup{log}\nu, H_1$ and $H_2$ for 200 model evaluations (left) and 500 model evaluations (right). Histograms represent 8092 values. True parameter values ($\theta = (\log 0.032, \log 9.5 \times 10^{-6}, 0.25, 0.85)$ for $\log \mu, \log \nu, H_1$ and $H_2$) are denoted by vertical black lines.}
    \label{fig:QGposterior}
\end{figure}

We begin by examining histograms of the posterior distributions for CES, HM, BOED, and GBOED using 200 and 500 model evaluations. Distributions for $\mu$ and $\nu$ are shown in log space. For both budgets, the posterior distributions produced by all methods include the true parameter values (Figure \ref{fig:QGposterior}, left and right panels).

Increasing the number of model evaluations from 200 to 500 generally leads to more concentrated and unimodal posteriors, most notably for BOED (blue). For parameter $\nu$, the posterior mass is more tightly concentrated around the true value for HM and BOED, whereas CES and GBOED place relatively less posterior mass near the true value for both evaluation budgets. For parameter $H_2$, increasing the number of model evaluations appears to slightly broaden the HM posterior, shifting mass away from the true value. Overall, increasing the number of model evaluations does not consistently result in substantial changes to posterior shape across methods, a behavior that mirrors what we observed for the Lorenz ’96 system.

\begin{figure}[ht]
    \centering
    \includegraphics[scale=.65]{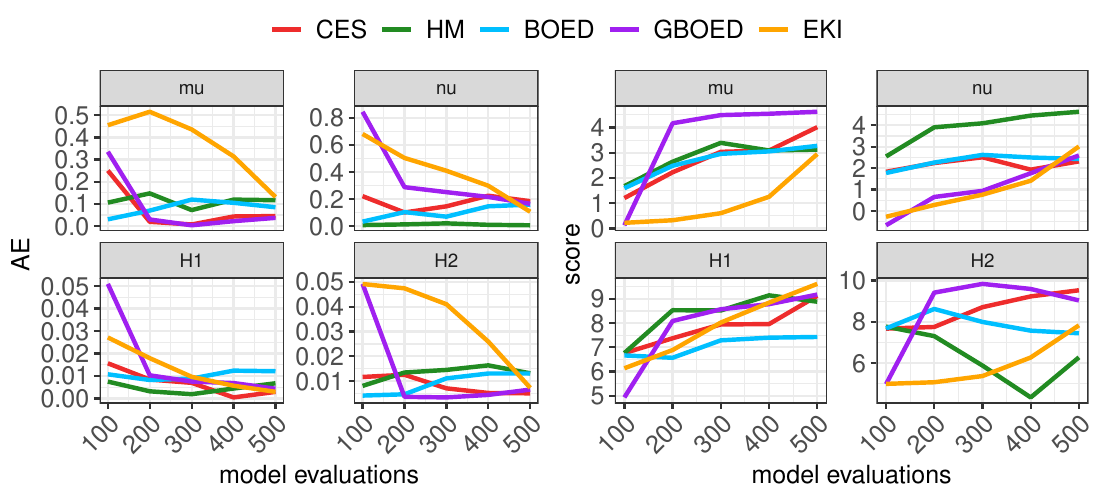}
    \caption{AEs (left) and scores (right) for CES, EKI, HM, BOED, and GBOED over
increasing QG model evaluations, based on true parameters $\theta = (0.032, 9.5 \times 10^{-6}, 0.25, 0.85)$ for $\mu, \nu, H_1$ and $H_2$. Results for $\mu$ and $\nu$ are computed in log space.}
    \label{fig:QGconvergence}
\end{figure}

We assessed convergence by evaluating absolute error (AE) and score every 100 model evaluations, from 100 to 500 evaluations, for all methods. In contrast to the convergence behavior observed for the Lorenz ’96 multiscale system, most methods exhibited clearer improvement trends for the QG system, particularly in terms of score (Figure \ref{fig:QGconvergence}, right).

For AE, EKI showed the most pronounced improvement as the number of model evaluations increased, with substantial reductions between 100 and 500 evaluations across all parameters (Figure \ref{fig:QGconvergence}, left). CES and GBOED exhibited rapid initial improvements from 100 to 200 evaluations, followed by a plateau, suggesting diminishing returns once the calibration-relevant region was sufficiently explored. In contrast, HM and BOED achieved relatively low AE values early on and displayed comparatively little variation as additional model evaluations were added.

Score-based convergence displayed an even clearer trend. Scores for nearly all methods and parameters increased steadily with additional model evaluations, indicating improved uncertainty quantification as posterior variance decreased (Figure \ref{fig:QGconvergence}, right). Overall, these results suggest that increasing model evaluations for the QG system leads to more consistent gains in both accuracy and uncertainty quantification than observed for the Lorenz ’96 multiscale system. Overall, the increase in performance exhibited by most methods for the QG system likely reflects stronger parameter identifiability and smoother parameter–statistic relationships than in the Lorenz ’96 multiscale system. In this setting, additional model evaluations may more reliably improve emulator fidelity in calibration-relevant regions, leading to gains in uncertainty quantification and, to a lesser extent, point estimation accuracy.

\section{Discussion and conclusions}
\label{sec:conclusions}
In this study, we compared four emulator-based Bayesian calibration strategies, BOED, GBOED, History Matching (HM), and Calibrate–Emulate–Sample (CES), alongside ensemble Kalman–based approaches and baseline LHS sampling, across two chaotic dynamical systems with very different computational and structural characteristics. Our results highlight both the robustness of Bayesian calibration methods and the importance of aligning design strategies with the downstream inferential task.

For the Lorenz ’96 multiscale and quasi-geostrophic (QG) systems, CES and its calibration component EKS or EKI consistently delivered strong performance in both accuracy and uncertainty quantification. When computational budgets were limited, CES provided clear advantages over EKS alone; however, these gains diminished as the number of model evaluations increased. GBOED and HM achieved performance comparable to CES and EKS even at relatively low model evaluation counts, suggesting that explicitly targeting calibration-relevant regions of parameter space, either through goal-oriented design or implausibility-based pruning, can be as effective as integrated, ensemble-based approaches.

In contrast, standard BOED generally underperformed relative to the other methods for the Lorenz ’96 multiscale system. This behavior likely reflects a mismatch between BOED’s objective to maximize information gain about the emulator and the needs of calibration, where accuracy in posterior-relevant regions is paramount. In chaotic systems with localized regions of posterior mass, GP emulators may generalize poorly outside well-sampled regions, making globally informative designs suboptimal for inference. However, this observation provides the impetus for developing goal-oriented extensions such as our GBOED algorithm, which explicitly aligns design decisions with posterior uncertainty reduction.

Although we did not perform the same extensive replicate-based analysis for the QG system as we did for Lorenz ’96 due to computational cost, all methods produced reasonable posterior estimates even with moderate evaluation budgets. Differences between methods were less pronounced, and no single approach clearly dominated. This convergence in performance likely reflects the smoother statistical dependence of QG diagnostics on parameters, which reduces sensitivity to design choices and mitigates the risks associated with emulator extrapolation. As a result, additional model evaluations tend to provide genuinely informative refinements to the emulator in calibration-relevant regions, leading to more consistent improvements in posterior concentration and score as computational budget increases. In this setting, EKI proved more stable than EKS, possibly due to reduced stochasticity in the update step combined with mild parameter confounding.

Taken together, these results suggest that for moderately smooth, high-cost systems such as QG, multiple emulator-based calibration strategies can yield comparable performance, and method selection may be guided more by practical considerations such as implementation complexity, parallelizability, and computational robustness than by accuracy alone. For more irregular or strongly nonlinear systems, however, calibration performance benefits from concentrating computational effort in regions of parameter space that are relevant for inference (GBOED, HM), or implicitly through iterative calibration steps (CES) leading to clear advantages over standard BOED. Recent results by \cite{reichelt2025calibration} suggest that active learning and BOED-based calibration methods yield the greatest benefit when observations strongly constrain the parameter space. This perspective helps explain why differences among CES, HM, and GBOED are relatively small for the QG system, while performance disparities are more pronounced for the Lorenz ’96 multiscale system, where the inverse problem may be less tightly constrained by the chosen summary statistics.

This work contributes to the limited literature on BOED-based calibration for chaotic and computationally intensive models \cite{dunbar2022ensemble, reichelt2025calibration}. While standard BOED may be ill-suited to some calibration problems, goal-oriented design strategies appear promising and may be further strengthened by integration with ensemble-based frameworks such as CES. Future work should explore hybrid approaches that combine the global efficiency of ensemble methods with the targeted refinement of goal-oriented design, as well as extend these comparisons to higher-dimensional parameter spaces and alternative emulator architectures.
Additionally, while batching was essential for practical deployment of BOED and GBOED on HPC systems, the choice of batch size was driven primarily by queue constraints rather than methodological considerations. The broader question of how batch size and number of iterations should be balanced under a fixed model evaluation budget and how these choices impact convergence and posterior quality remains an important topic for future study.

\section*{Data and code availability}
All code and data used for this study is available in a Zenodo repository:\\ https://zenodo.org/records/13761230.

\section*{Acknowledgments}
We would like to thank early reviewers for improving our manuscript.

\bibliographystyle{siamplain}
\bibliography{refs}

@article{cleary2021calibrate,
  title={Calibrate, emulate, sample},
  author={Cleary, Emmet and Garbuno-Inigo, Alfredo and Lan, Shiwei and Schneider, Tapio and Stuart, Andrew M},
  journal={Journal of Computational Physics},
  volume={424},
  pages={109716},
  year={2021},
  publisher={Elsevier}
}

@article{lguensat2023semi,
  title={Semi-automatic tuning of coupled climate models with multiple intrinsic timescales: Lessons learned from the {L}orenz '96 model},
  author={Lguensat, Redouane and Deshayes, Julie and Durand, Homer and Balaji, Venkatramani},
  journal={Journal of Advances in Modeling Earth Systems},
  volume={15},
  number={5},
  pages={e2022MS003367},
  year={2023},
  publisher={Wiley Online Library}
}

@article{mckay+bc79,
 ISSN = {00401706},
 URL = {http://www.jstor.org/stable/1268522},
 author = {M. D. McKay and R. J. Beckman and W. J. Conover},
 journal = {Technometrics},
 number = {2},
 pages = {239--245},
 publisher = {[Taylor & Francis, Ltd., American Statistical Association, American Society for Quality]},
 title = {A Comparison of Three Methods for Selecting Values of Input Variables in the Analysis of Output from a Computer Code},
 urldate = {2024-10-16},
 volume = {21},
 year = {1979}
}

@article{lindley1956measure,
  title={On a measure of the information provided by an experiment},
  author={Lindley, Dennis V},
  journal={The Annals of Mathematical Statistics},
  volume={27},
  number={4},
  pages={986--1005},
  year={1956},
  publisher={Institute of Mathematical Statistics}
}

@article{schillings2017analysis,
  title={Analysis of the ensemble Kalman filter for inverse problems},
  author={Schillings, Claudia and Stuart, Andrew M},
  journal={SIAM Journal on Numerical Analysis},
  volume={55},
  number={3},
  pages={1264--1290},
  year={2017},
  publisher={SIAM}
}

@article{wu2021efficient,
  title={An efficient method for goal-oriented linear {B}ayesian optimal experimental design: Application to optimal sensor placement},
  author={Wu, Keyi and Chen, Peng and Ghattas, Omar},
  journal={arXiv preprint arXiv:2102.06627},
  year={2021}
}

@article{zhong2024goal,
  title={Goal-Oriented {B}ayesian Optimal Experimental Design for Nonlinear Models using {M}arkov Chain {M}onte {C}arlo},
  author={Zhong, Shijie and Shen, Wanggang and Catanach, Tommie and Huan, Xun},
  journal={arXiv preprint arXiv:2403.18072},
  year={2024}
}

@article{attia2018goal,
  title={Goal-oriented optimal design of experiments for large-scale {B}ayesian linear inverse problems},
  author={Attia, Ahmed and Alexanderian, Alen and Saibaba, Arvind K},
  journal={Inverse Problems},
  volume={34},
  number={9},
  pages={095009},
  year={2018},
  publisher={IOP Publishing}
}

@article{bernardo1979expected,
  title={Expected information as expected utility},
  author={Bernardo, Jos{\'e} M},
  journal={the Annals of Statistics},
  pages={686--690},
  year={1979},
  publisher={JSTOR}
}

@inproceedings{ferrolino2020optimal,
  title={Optimal location of sensors for early detection of tsunami waves},
  author={Ferrolino, Angelie R and Lope, Jose Ernie C and Mendoza, Renier G},
  booktitle={Computational Science--ICCS 2020: 20th International Conference, Amsterdam, The Netherlands, June 3--5, 2020, Proceedings, Part II 20},
  pages={562--575},
  year={2020},
  organization={Springer}
}

@article{ginebra2007measure,
  title={On the ieasure of the information in a statistical experiment},
  author={Ginebra, Josep},
  journal={Bayesian Analysis},
  volume={2},
  number={1},
  pages={167--212},
  year={2007}
}

@inproceedings{lorenz1996predictability,
  title={Predictability: A problem partly solved},
  author={Lorenz, Edward N},
  booktitle={Proc. Seminar on predictability},
  volume={1},
  number={1},
  year={1996},
  organization={Reading}
}

@article{catanach2020bayesian,
  title={Bayesian inference of stochastic reaction networks using multifidelity sequential tempered {M}arkov {C}hain {M}onte {C}arlo},
  author={Catanach, Thomas A and Vo, Huy D and Munsky, Brian},
  journal={International journal for uncertainty quantification},
  volume={10},
  number={6},
  year={2020},
  publisher={Begel House Inc.}
}

@article{pukelsheim94,
 ISSN = {00031305},
 URL = {http://www.jstor.org/stable/2684253},
 author = {Friedrich Pukelsheim},
 journal = {The American Statistician},
 number = {2},
 pages = {88--91},
 publisher = {[American Statistical Association, Taylor & Francis, Ltd.]},
 title = {The three sigma rule},
 urldate = {2024-09-17},
 volume = {48},
 year = {1994}
}

@article{drovandi+np21,
author = {Drovandi, Christopher and Nott, David J. and Pagendam, Daniel E.},
title = {A semiautomatic method for history matching using sequential {M}onte {C}arlo},
journal = {SIAM/ASA Journal on Uncertainty Quantification},
volume = {9},
number = {3},
pages = {1034-1063},
year = {2021},
doi = {10.1137/19M1286694},

URL = { 
    
        https://doi.org/10.1137/19M1286694
    
    

},
eprint = { 
    
        https://doi.org/10.1137/19M1286694
    
    

}
}

@article{ehrett2021simultaneous,
  title={Simultaneous {B}ayesian calibration and engineering design with an application to a vibration isolation system},
  author={Ehrett, Carl and Brown, D Andrew and Kitchens, Christopher and Xu, Xinyue and Platz, Roland and Atamturktur, Sez},
  journal={Journal of Verification, Validation and Uncertainty Quantification},
  volume={6},
  number={1},
  pages={011007},
  year={2021},
  publisher={American Society of Mechanical Engineers}
}

@inproceedings{ling2013challenging,
  title={Challenging issues in {B}ayesian calibration of multi-physics models},
  author={Ling, You and Mahadevan, Sankaran},
  booktitle={54th AIAA/ASME/ASCE/AHS/ASC Structures, Structural Dynamics, and Materials Conference},
  pages={1874},
  year={2013}
}

@article{viana2021survey,
  title={A survey of {B}ayesian calibration and physics-informed neural networks in scientific modeling},
  author={Viana, Felipe AC and Subramaniyan, Arun K},
  journal={Archives of Computational Methods in Engineering},
  volume={28},
  number={5},
  pages={3801--3830},
  year={2021},
  publisher={Springer}
}

@article{durr2023bayesian,
  title={Bayesian calibration of {MEMS} accelerometers},
  author={D{\"u}rr, Oliver and Fan, Po-Yu and Yin, Zong-Xian},
  journal={IEEE Sensors Journal},
  volume={23},
  number={12},
  pages={13319--13326},
  year={2023},
  publisher={IEEE}
}

@article{helleckes2022bayesian,
  title={Bayesian calibration, process modeling and uncertainty quantification in biotechnology},
  author={Helleckes, Laura Marie and Osthege, Michael and Wiechert, Wolfgang and von Lieres, Eric and Oldiges, Marco},
  journal={PLoS computational biology},
  volume={18},
  number={3},
  pages={e1009223},
  year={2022},
  publisher={Public Library of Science San Francisco, CA USA}
}

@article{kennedy2001bayesian,
  title={Bayesian calibration of computer models},
  author={Kennedy, Marc C and O'Hagan, Anthony},
  journal={Journal of the Royal Statistical Society: Series B (Statistical Methodology)},
  volume={63},
  number={3},
  pages={425--464},
  year={2001},
  publisher={Wiley Online Library}
}

@article{yarger2024autocalibration,
  title={Autocalibration of the {E3SM} version 2 atmosphere model using a {PCA}-based surrogate for spatial fields},
  author={Yarger, Drew and Wagman, Benjamin Moore and Chowdhary, Kenny and Shand, Lyndsay},
  journal={Journal of Advances in Modeling Earth Systems},
  volume={16},
  number={4},
  pages={e2023MS003961},
  year={2024},
  publisher={Wiley Online Library}
}

@article{chang2014fast,
  title={Fast dimension-reduced climate model calibration and the effect of data aggregation},
  author={Chang, Won and Haran, Murali and Olson, Roman and Keller, Klaus},
  year={2014}
}

@article{dunbar2021calibration,
  title={Calibration and uncertainty quantification of convective parameters in an idealized {GCM}},
  author={Dunbar, Oliver RA and Garbuno-Inigo, Alfredo and Schneider, Tapio and Stuart, Andrew M},
  journal={Journal of Advances in Modeling Earth Systems},
  volume={13},
  number={9},
  pages={e2020MS002454},
  year={2021},
  publisher={Wiley Online Library}
}

@article{chaloner1995bayesian,
  title={Bayesian experimental design: A review},
  author={Chaloner, Kathryn and Verdinelli, Isabella},
  journal={Statistical science},
  pages={273--304},
  year={1995},
  publisher={JSTOR}
}

@article{owen2003quasi,
  title={Quasi-{M}onte {C}arlo {S}ampling},
  author={Owen, Art B},
  journal={Monte Carlo Ray Tracing: Siggraph},
  volume={1},
  pages={69--88},
  year={2003},
  publisher={Citeseer}
}

@article{jackson2008error,
  title={Error reduction and convergence in climate prediction},
  author={Jackson, Charles S and Sen, Mrinal K and Huerta, Gabriel and Deng, Yi and Bowman, Kenneth P},
  journal={Journal of Climate},
  volume={21},
  number={24},
  pages={6698--6709},
  year={2008},
  publisher={American Meteorological Society}
}

@article{smith2024fair,
  title={Fair-calibrate v1. 4.1: calibration, constraining, and validation of the {FaIR} simple climate model for reliable future climate projections},
  author={Smith, Chris and Cummins, Donald P and Fredriksen, Hege-Beate and Nicholls, Zebedee and Meinshausen, Malte and Allen, Myles and Jenkins, Stuart and Leach, Nicholas and Mathison, Camilla and Partanen, Antti-Ilari},
  journal={Geoscientific Model Development},
  volume={17},
  number={23},
  pages={8569--8592},
  year={2024},
  publisher={Copernicus Publications G{\"o}ttingen, Germany}
}

@article{tienstra2024early,
  title={Early stopping for ensemble {K}alman-{B}ucy inversion},
  author={Tienstra, Maia and Reich, Sebastian},
  journal={arXiv preprint arXiv:2403.18353},
  year={2024}
}

@article{huang2022iterated,
  title={Iterated Kalman methodology for inverse problems},
  author={Huang, Daniel Zhengyu and Schneider, Tapio and Stuart, Andrew M},
  journal={Journal of Computational Physics},
  volume={463},
  pages={111262},
  year={2022},
  publisher={Elsevier}
}

@article{chada2020iterative,
  title={Iterative ensemble {K}alman methods: A unified perspective with some new variants},
  author={Chada, Neil K and Chen, Yuming and Sanz-Alonso, Daniel},
  journal={arXiv preprint arXiv:2010.13299},
  year={2020}
}

@article{chada2020tikhonov,
  title={Tikhonov regularization within ensemble {K}alman inversion},
  author={Chada, Neil K and Stuart, Andrew M and Tong, Xin T},
  journal={SIAM Journal on Numerical Analysis},
  volume={58},
  number={2},
  pages={1263--1294},
  year={2020},
  publisher={SIAM}
}

@software{climateAlliance,
  author = {{Climate Alliance}},
  title = {EnsembleKalmanProcesses},
  url = {https://clima.github.io/EnsembleKalmanProcesses.jl/dev/ensemble_kalman_sampler/},
  version = {2.5.0},
  date = {2025-07-30},
}

@article{parthipan2023using,
  title={Using probabilistic machine learning to better model temporal patterns in parameterizations: a case study with the {L}orenz 96 model},
  author={Parthipan, Raghul and Christensen, Hannah M and Hosking, J Scott and Wischik, Damon J},
  journal={Geoscientific Model Development},
  volume={16},
  number={15},
  pages={4501--4519},
  year={2023},
  publisher={Copernicus Publications G{\"o}ttingen, Germany}
}

@article{orrell2003model,
  title={Model error and predictability over different timescales in the {L}orenz'96 systems},
  author={Orrell, David},
  journal={Journal of the atmospheric sciences},
  volume={60},
  number={17},
  pages={2219--2228},
  year={2003},
  publisher={American Meteorological Society}
}

@article{ott2004local,
  title={A local ensemble {K}alman filter for atmospheric data assimilation},
  author={Ott, Edward and Hunt, Brian R and Szunyogh, Istvan and Zimin, Aleksey V and Kostelich, Eric J and Corazza, Matteo and Kalnay, Eugenia and Patil, DJ and Yorke, James A},
  journal={Tellus A: Dynamic Meteorology and Oceanography},
  volume={56},
  number={5},
  pages={415--428},
  year={2004},
  publisher={Taylor \& Francis}
}

@book{gramacy2020surrogates,
  title={Surrogates: {G}aussian process modeling, design, and optimization for the applied sciences},
  author={Gramacy, Robert B},
  year={2020},
  publisher={Chapman and Hall/CRC}
}

@article{novak2018polynomial,
  title={Polynomial chaos expansion for surrogate modelling: theory and software},
  author={Novak, Lukas and Novak, Drahomir},
  journal={Beton-und Stahlbetonbau},
  volume={113},
  pages={27--32},
  year={2018},
  publisher={Wiley Online Library}
}

@article{gneiting2002compactly,
  title={Compactly supported correlation functions},
  author={Gneiting, Tilmann},
  journal={Journal of Multivariate Analysis},
  volume={83},
  number={2},
  pages={493--508},
  year={2002},
  publisher={Elsevier}
}

@article{katzfuss2021general,
  title={A general framework for {V}ecchia approximations of {G}aussian processes},
  author={Katzfuss, Matthias and Guinness, Joseph},
  journal={Statistical Science},
  volume={36},
  number={1},
  pages={124--141},
  year={2021}
}

@article{katzfuss2020vecchia,
  title={Vecchia approximations of {G}aussian-process predictions},
  author={Katzfuss, Matthias and Guinness, Joseph and Gong, Wenlong and Zilber, Daniel},
  journal={Journal of Agricultural, Biological and Environmental Statistics},
  volume={25},
  pages={383--414},
  year={2020},
  publisher={Springer}
}

@article{constantinou2021geophysicalflows,
  title={GeophysicalFlows. jl: Solvers for geophysical fluid dynamics problems in periodic domains on CPUs {GPUs}},
  author={Constantinou, Navid and Wagner, Gregory and Siegelman, Lia and Pearson, Brodie and Pal{\'o}czy, Andr{\'e}},
  journal={Journal of Open Source Software},
  volume={6},
  number={60},
  year={2021}
}

@article{vecchia1988estimation,
  title={Estimation and model identification for continuous spatial processes},
  author={Vecchia, Aldo V},
  journal={Journal of the Royal Statistical Society: Series B (Methodological)},
  volume={50},
  number={2},
  pages={297--312},
  year={1988},
  publisher={Wiley Online Library}
}

@article{zammit2025neural,
  title={Neural methods for amortized inference},
  author={Zammit-Mangion, Andrew and Sainsbury-Dale, Matthew and Huser, Rapha{\"e}l},
  journal={Annual Review of Statistics and Its Application},
  volume={12},
  number={1},
  pages={311--335},
  year={2025},
  publisher={Annual Reviews}
}

@article{cao2022scalable,
  title={Scalable {G}aussian-process regression and variable selection using {V}ecchia approximations},
  author={Cao, Jian and Guinness, Joseph and Genton, Marc G and Katzfuss, Matthias},
  journal={Journal of Machine Learning Research},
  volume={23},
  number={348},
  pages={1--30},
  year={2022}
}

@article{datta2016hierarchical,
  title={Hierarchical nearest-neighbor {G}aussian process models for large geostatistical datasets},
  author={Datta, Abhirup and Banerjee, Sudipto and Finley, Andrew O and Gelfand, Alan E},
  journal={Journal of the American Statistical Association},
  volume={111},
  number={514},
  pages={800--812},
  year={2016},
  publisher={Taylor \& Francis}
}

@article{gramacy2016lagp,
  title={la{GP}: large-scale spatial modeling via local approximate {G}aussian processes in {R}},
  author={Gramacy, Robert B},
  journal={Journal of Statistical Software},
  volume={72},
  pages={1--46}
}

@article{emery2009kriging,
  title={The kriging update equations and their application to the selection of neighboring data},
  author={Emery, X},
  journal={Computational Geosciences}, 
  volume={13},
  number={3},
  pages={269--280},
  year={2009},
  publisher={Springer}
}

@article{hensman2013gaussian,
  title={{G}aussian processes for big data},
  author={Hensman, James and Fusi, Nicolo and Lawrence, Neil D},
  journal={arXiv preprint arXiv:1309.6835},
  year={2013}
}

@article{springer2021efficient,
  title={Efficient Bayesian inference for large chaotic dynamical systems},
  author={Springer, Sebastian and Haario, Heikki and Susiluoto, Jouni and Bibov, Aleksandr and Davis, Andrew and Marzouk, Youssef},
  journal={Geoscientific Model Development},
  volume={14},
  number={7},
  pages={4319--4333},
  year={2021},
  publisher={Copernicus Publications G{\"o}ttingen, Germany}
}

@article{cressie2008fixed,
  title={Fixed rank kriging for very large spatial data sets},
  author={Cressie, Noel and Johannesson, Gardar},
  journal={Journal of the Royal Statistical Society: Series B (Statistical Methodology)},
  volume={70},
  number={1},
  pages={209--226},
  year={2008},
  publisher={Wiley Online Library}
}

@article{banerjee2008gaussian,
  title={{G}aussian predictive process models for large spatial data sets},
  author={Banerjee, Sudipto and Gelfand, Alan E and Finley, Andrew O and Sang, Huiyan},
  journal={Journal of the Royal Statistical Society: Series B (Statistical Methodology)},
  volume={70},
  number={4},
  pages={825--848},
  year={2008},
  publisher={Wiley Online Library}
}

@article{quinonero2005unifying,
  title={A unifying view of sparse approximate {G}aussian process regression},
  author={Quinonero-Candela, Joaquin and Rasmussen, Carl Edward},
  journal={The Journal of Machine Learning Research},
  volume={6},
  pages={1939--1959},
  year={2005},
  publisher={JMLR. org}
}

@article{kaufman2011efficient,
  title={Efficient emulators of computer experiments using compactly supported correlation functions, with an application to cosmology},
  author={Kaufman, Cari G and Bingham, Derek and Habib, Salman and Heitmann, Katrin and Frieman, Joshua A},
  journal = {Annals of Applied Statistics},
  volume={5},
  number={4},
  pages = {2470--2492},
  year={2011}
}

@article{dunbar2022ensemble,
  title={Ensemble-based experimental design for targeting data acquisition to inform climate models},
  author={Dunbar, Oliver RA and Howland, Michael F and Schneider, Tapio and Stuart, Andrew M},
  journal={Journal of Advances in Modeling Earth Systems},
  volume={14},
  number={9},
  pages={e2022MS002997},
  year={2022},
  publisher={Wiley Online Library}
}

@Manual{baseR,
    title = {R: A language and environment for statistical computing},
    author = {{R Core Team}},
    organization = {R Foundation for Statistical Computing},
    address = {Vienna, Austria},
    year = {2023},
    url = {https://www.R-project.org/},
}

@Article{modelbasedPackage,
    title = {Estimation of model-based predictions, contrasts and means.},
    author = {Dominique Makowski and Mattan S. Ben-Shachar and Indrajeet Patil and Daniel Lüdecke},
    journal = {CRAN},
    year = {2020},
    url = {https://github.com/easystats/modelbased},
}

@Manual{emmeansPackage,
    title = {emmeans: estimated marginal means, aka least-squares means},
    author = {Russell V. Lenth},
    year = {2024},
    note = {R package version 1.10.5},
    url = {https://CRAN.R-project.org/package=emmeans},
}

@article{ricciardi2024bayesian,
  title={Bayesian optimal experimental design for constitutive model calibration},
  author={Ricciardi, Denielle E and Seidl, Daniel Thomas and Lester, Brian T and Jones, Amanda Rose and Jones, EMC},
  journal={International Journal of Mechanical Sciences},
  volume={265},
  pages={108881},
  year={2024},
  publisher={Elsevier}
}

@article{gneiting2007strictly,
  title={Strictly proper scoring rules, prediction, and estimation},
  author={Gneiting, Tilmann and Raftery, Adrian E},
  journal={Journal of the American statistical Association},
  volume={102},
  number={477},
  pages={359--378},
  year={2007},
  publisher={Taylor \& Francis}
}

@article{king2024bayesian,
  title={Bayesian history matching applied to the calibration of a gravity wave parameterization},
  author={King, Robert C and Mansfield, Laura A and Sheshadri, Aditi},
  journal={Journal of Advances in Modeling Earth Systems},
  volume={16},
  number={4},
  pages={e2023MS004163},
  year={2024},
  publisher={Wiley Online Library}
}

@article{edwards2011precalibrating,
  title={Precalibrating an intermediate complexity climate model},
  author={Edwards, Neil R and Cameron, David and Rougier, Jonathan},
  journal={Climate dynamics},
  volume={37},
  pages={1469--1482},
  year={2011},
  publisher={Springer}
}

@incollection{craig1997pressure,
  title={Pressure matching for hydrocarbon reservoirs: a case study in the use of {B}ayes linear strategies for large computer experiments},
  author={Craig, Peter S and Goldstein, Michael and Seheult, Allan H and Smith, James A},
  booktitle={Case Studies in Bayesian Statistics: Volume III},
  pages={37--93},
  year={1997},
  publisher={Springer}
}

@article{raoult2024exploring,
  title={Exploring the potential of history matching for land surface model calibration},
  author={Raoult, Nina and Beylat, Simon and Salter, James M and Hourdin, Fr{\'e}d{\'e}ric and Bastrikov, Vladislav and Ottl{\'e}, Catherine and Peylin, Philippe},
  journal={Geoscientific Model Development},
  volume={17},
  number={15},
  pages={5779--5801},
  year={2024},
  publisher={Copernicus Publications G{\"o}ttingen, Germany}
}

@article{williamson2013history,
  title={History matching for exploring and reducing climate model parameter space using observations and a large perturbed physics ensemble},
  author={Williamson, Daniel and Goldstein, Michael and Allison, Lesley and Blaker, Adam and Challenor, Peter and Jackson, Laura and Yamazaki, Kuniko},
  journal={Climate dynamics},
  volume={41},
  pages={1703--1729},
  year={2013},
  publisher={Springer}
}

@article{couvreux2021process,
  title={Process-based climate model development harnessing machine learning: I. A calibration tool for parameterization improvement},
  author={Couvreux, Fleur and Hourdin, Fr{\'e}d{\'e}ric and Williamson, Daniel and Roehrig, Romain and Volodina, Victoria and Villefranque, Najda and Rio, Catherine and Audouin, Olivier and Salter, James and Bazile, Eric and others},
  journal={Journal of Advances in Modeling Earth Systems},
  volume={13},
  number={3},
  pages={e2020MS002217},
  year={2021},
  publisher={Wiley Online Library}
}

@article{williamson2015identifying,
  title={Identifying and removing structural biases in climate models with history matching},
  author={Williamson, Daniel and Blaker, Adam T and Hampton, Charlotte and Salter, James},
  journal={Climate dynamics},
  volume={45},
  pages={1299--1324},
  year={2015},
  publisher={Springer}
}

@article{garbuno2020interacting,
  title={Interacting {L}angevin diffusions: gradient structure and ensemble {K}alman sampler},
  author={Garbuno-Inigo, Alfredo and Hoffmann, Franca and Li, Wuchen and Stuart, Andrew M},
  journal={SIAM Journal on Applied Dynamical Systems},
  volume={19},
  number={1},
  pages={412--441},
  year={2020},
  publisher={SIAM}
}

@book{williams2006gaussian,
  title={Gaussian processes for machine learning},
  author={Williams, Christopher KI and Rasmussen, Carl Edward},
  volume={2},
  number={3},
  year={2006},
  publisher={MIT press Cambridge, MA}
}

@article{kovachki2019ensemble,
  title={Ensemble {K}alman inversion: a derivative-free technique for machine learning tasks},
  author={Kovachki, Nikola B and Stuart, Andrew M},
  journal={Inverse Problems},
  volume={35},
  number={9},
  pages={095005},
  year={2019},
  publisher={IOP Publishing}
}

@article{fandry1984two,
  title={A two-layer quasi-geostrophic model of summer trough formation in the {A}ustralian subtropical easterlies},
  author={Fandry, CB and Leslie, LM},
  journal={Journal of Atmospheric Sciences},
  volume={41},
  number={5},
  pages={807--818},
  year={1984}
}

@article{wilkinson2010bayesian,
  title={Bayesian calibration of expensive multivariate computer experiments},
  author={Wilkinson, Richard D},
  journal={Large-scale inverse problems and quantification of uncertainty},
  pages={195--215},
  year={2010},
  publisher={Wiley Online Library}
}

@article{reichelt2025calibration,
  title={Calibration of Climate Model Parameterizations using {B}ayesian Experimental Design},
  author={Reichelt, Tim and Rainforth, Tom and Watson-Parris, Duncan},
  journal={Machine Learning: Earth},
  year={2025}
}

@article{iglesias2013ensemble,
  title={Ensemble Kalman methods for inverse problems},
  author={Iglesias, Marco A and Law, Kody JH and Stuart, Andrew M},
  journal={Inverse Problems},
  volume={29},
  number={4},
  pages={045001},
  year={2013},
  publisher={IOP Publishing}
}

@ARTICLE{bocquet23,
  
AUTHOR={Bocquet, Marc},   
	 
TITLE={{Surrogate modeling for the climate sciences dynamics with machine learning and data assimilation}},      
	
JOURNAL={Frontiers in Applied Mathematics and Statistics},      
	
VOLUME={9},           
	
YEAR={2023},      
	  
URL={https://www.frontiersin.org/articles/10.3389/fams.2023.1133226},       
	
DOI={10.3389/fams.2023.1133226},      
	
ISSN={2297-4687}
}

@Article{weber+chkl20,
AUTHOR = {Weber, T. and Corotan, A. and Hutchinson, B. and Kravitz, B. and Link, R.},
TITLE = {{Technical note: Deep learning for creating surrogate models of
precipitation in Earth system models}},
JOURNAL = {Atmospheric Chemistry and Physics},
VOLUME = {20},
YEAR = {2020},
NUMBER = {4},
PAGES = {2303--2317},
URL = {https://acp.copernicus.org/articles/20/2303/2020/},
DOI = {10.5194/acp-20-2303-2020}
}

@article{lutsko+cgmm21,
title = {Machine Learning for Surrogate Modeling of the Upper Ocean and Heat Exchange Between the Ocean and Atmosphere},
author = {Lutsko, Nicholas and Cornuelle, Bruce and Gille, Sarah and Mazloff, Mathew and Morzfeld, Matthias},
doi = {10.2172/1769742},
url = {https://www.osti.gov/biblio/1769742}, 
place = {United States},
year = {2021},
month = {4}
}

@article{howland+ds22,
author = {Howland, Michael F. and Dunbar, Oliver R. A. and Schneider, Tapio},
title = {{Parameter Uncertainty Quantification in an Idealized {GCM} With a Seasonal Cycle}},
journal = {Journal of Advances in Modeling Earth Systems},
volume = {14},
number = {3},
pages = {e2021MS002735},
doi = {https://doi.org/10.1029/2021MS002735},
url = {https://agupubs.onlinelibrary.wiley.com/doi/abs/10.1029/2021MS002735},
eprint = {https://agupubs.onlinelibrary.wiley.com/doi/pdf/10.1029/2021MS002735},
note = {e2021MS002735 2021MS002735} ,
year = {2022}
}

\end{document}